\documentclass[article]{siamart0516}

\usepackage{amsmath, amsfonts}
\usepackage{subfig}
\usepackage{comment}
\usepackage{graphicx,color}
\usepackage[]{cite}
\usepackage{algpseudocode,algorithm}
\usepackage{placeins}

\newtheorem{remark}{Remark}

\renewcommand{\eqref}[1]{(\ref{#1})}

\newcommand{\bvec}[1]{\mathbf{#1}}

\newcommand{\mc}[1]{\mathcal{#1}}

\newcommand{\va}{\bvec{a}}

\newcommand{\vr}{\bvec{r}}
\newcommand{\vk}{\bvec{k}}
\newcommand{\vR}{\bvec{R}}

\newcommand{\ud}{\,\mathrm{d}}

\newcommand{\abs}[1]{\lvert#1\rvert}

\newcommand{\bra}[1]{\langle#1\rvert}
\newcommand{\ket}[1]{\lvert#1\rangle}
\newcommand{\wt}[1]{\widetilde{#1}}

\newcommand{\ie}{\emph{i.e.}}
\newcommand{\eg}{\emph{e.g.}}

\newcommand{\RR}{\mathbb{R}}
\newcommand{\CC}{\mathbb{C}}
\newcommand{\ZZ}{\mathbb{Z}}
\newcommand{\I}{\imath}

\newcommand{\ang}{\text{\AA}} 

\DeclareMathOperator*{\argmax}{arg\,max}

\newcommand{\response}[1]{{#1}}

\title{Disentanglement via entanglement: A unified method for Wannier localization}

\author{
  Anil Damle
  \thanks{Department of Computer Science, Cornell University, Ithaca, NY 14853 (\email{damle@cornell.edu}). \funding{National Science Foundation Mathematical Sciences
Postdoctoral Research Fellowship (grant number DMS-1606277).}}
\and
Lin Lin
  \thanks{Department of Mathematics, University of California, Berkeley,
  Berkeley, CA 94720; and Computational Research Division, Lawrence
  Berkeley National Laboratory, Berkeley, CA 94720
  (\email{linlin@math.berkeley.edu}). \funding{U.S. Department of Energy
  (contract number DE-SC0017867, the DOE Center for
  Applied Mathematics for Energy Research Applications (CAMERA)
  program),  National Science Foundation (grant number DMS-1652330) and
  Alfred P. Sloan fellowship.}}
}

\headers{A unified method for Wannier localization}{Anil Damle and Lin Lin}
\begin{document}

\maketitle

\begin{abstract}
The Wannier localization problem in quantum physics is mathematically analogous to finding a localized representation of a subspace corresponding to a nonlinear eigenvalue problem.  While Wannier localization is well understood for insulating materials with isolated eigenvalues, less is known for metallic systems with entangled eigenvalues. Currently, the most widely used method for treating systems with entangled eigenvalues is to first obtain a reduced subspace (often referred to as disentanglement) and then to solve the Wannier localization problem by treating the reduced subspace as an isolated system. This is a multi-objective nonconvex optimization procedure and its solution can depend sensitively on the initial guess.  We propose a new method to solve the Wannier localization problem, avoiding the explicit use of an an optimization procedure.  Our method is robust, efficient, relies on few tunable parameters, and provides a unified framework for addressing problems with isolated and entangled eigenvalues. 
\end{abstract}



% \ra{1.4}

\section{Introduction}
\label{sec:intro}

Localized representations of electronic wavefunctions have a
wide range of applications in quantum physics, chemistry, and
materials science.  They require
significantly less memory to store, and are the foundation of
so-called ``linear scaling
methods''~\cite{Kohn1996,Goedecker1999,BowlerMiyazaki2012}
for solving quantum problems.  They can also be used to analyze the chemical
bonding in complex materials, interpolate the band structure of
crystals, accelerate ground and excited state electronic structure
calculations, and form reduced order models for strongly correlated
many body systems~\cite{MarzariMostofiYatesEtAl2012}.

In an effective single particle theory
such as the Kohn-Sham density
functional theory (KSDFT)~\cite{HohenbergKohn1964,KohnSham1965}, the electronic wavefunctions are
given by the (possibly generalized) eigenfunctions, denoted $\{\psi_{i}(\vr)\}$, of a self-adjoint Hamiltonian operator
$\mc{H}$:
\begin{equation}
  \label{eqn:setup}
  \mc{H}\psi_{i}(\vr) = \varepsilon_{i} \psi_{i}(\vr),\quad \varepsilon_{i}\in
  \mc{I}.
\end{equation}
Here $\mc{I}$ is an interval that can be interpreted as an energy window that indicates the eigenfunctions of physical interest. 
These eigenfunctions are generally delocalized, \ie~have significant magnitude in large portions of the computational
domain. 
The Wannier localization problem is as follows:
find an approximately minimal set of
orthonormal and localized
functions $\{w_{j}\}$, which have significant magnitude on only a small portion of the computational domain, such that
\[
\text{span}\{\psi_{j}\}_{\varepsilon_{i}\in \mc{I}} \subseteq \mathcal{V}_{w} \mathrel{\mathop:}= \text{span}\{w_{j}\}.
\]
Following the convention in quantum physics, $\{w_{j}\}$ are called
Wannier functions~\cite{Wannier1937,Kohn1959,Blount1962}. 

When the eigenvalues in $\mc{I}$  are \emph{isolated}, \ie~
\begin{equation}
  \label{eqn:isolation}
\inf_{\varepsilon_{i}\in \mc{I},\varepsilon_{i'}\notin\mc{I}}
\abs{\varepsilon_{i}-\varepsilon_{i'}}>0,
\end{equation}
the Wannier localization problem has been studied extensively in
mathematics and
physics~\cite{Kohn1959,Blount1962,Nenciu1991,MarzariVanderbilt1997,KochGoedecker2001,BrouderPanatiCalandraEtAl2007,Gygi2009,ELiLu2010,OzolinsLaiCaflischEtAl2013,PanatiPisante2013,DamleLinYing2015,MustafaCohCohenEtAl2015,CancesLevittPanatiEtAl2017,DamleLinYing2017,Cornean2017WannierTRS,Cornean2017wanniermetalic}. 
Loosely speaking, for a class of Hamiltonians $\mc{H}$, one can
construct exponentially localized Wannier functions such that
$\mathcal{V}_{w} = \text{span}\{\psi_{i}\}_{\varepsilon_{i}\in
\mc{I}}$. The isolation condition is satisfied when treating valence
bands of insulating systems.

When the isolation condition~\eqref{eqn:isolation} is
violated, the eigenvalues in $\mc{I}$ become \emph{entangled}. Entangled
eigenvalues appear ubiquitously in metallic systems, but also insulating
systems when conduction bands or a selected range of valence bands are considered. The problem now becomes significantly more
difficult: both identify a subspace
$\mathcal{V}_{w}$ that admits a localized basis, and construct such a basis.

The most widely used method to construct localized functions in this
scenario is a \emph{disentanglement}
procedure~\cite{SouzaMarzariVanderbilt2001}. It first identifies
$\mathcal{V}_{w}$ by minimizing a nonlinear ``smoothness functional''.
Then it computes $\{w_{i}\}$ by minimizing a nonlinear ``spread
functional''~\cite{MarzariVanderbilt1997}. In both problems the feasible set is nonconvex. While this
two step method has been successfully applied to a number of
applications~\cite{MarzariMostofiYatesEtAl2012}, there is little
mathematical understanding of the disentanglement procedure. Sensitive
dependence on the initial guess along with a number of tunable
parameters in the optimization formulation gives rise to a number of practical difficulties in using this method. Often, detailed knowledge of the underlying physical system is required to obtain physically meaningful results.

In this manuscript, we propose a unified method
to address the Wannier localization problem for both isolated and
entangled cases.
Instead of an initial ``disentanglement'' step, our method explicitly constructs a
quasi-density matrix that ``entangles'' the eigenfunctions of interest
with the rest of the eigenfunctions in a controlled manner. 
This has the effect of simultaneously identifying the subspace $\mathcal{V}_{w}$ and
constructing the localized basis. For the isolated case our new method
reduces to the prior selected columns of the density matrix
(SCDM) method~\cite{DamleLinYing2015}, and hence we still refer to our new, unified
approach as SCDM. 

\response{The core technical contributions of this manuscript are the
extension of SCDM to the entangled band case through the use of a
quasi-density matrix, and a significantly simplified extension to
crystal systems than that of prior work~\cite{DamleLinYing2017}. 
While our methodology can be used to treat a broad range of physical
systems, the SCDM algorithm for the isolated case will
fail for topological insulators. In this setting, the band structure is
isolated but the corresponding isolated bands do not admit
well-localized Wannier functions due to topological
obstruction~\cite{BrouderPanatiCalandraEtAl2007,Cornean2017WannierTRS,Soluyanov2011WannierZ2}.}

The SCDM method has several significant advantages. First is its
simplicity. There are \response{essentially} no tunable parameters for the isolated case and only two parameters in the entangled case. Second, SCDM is constructed using
standard linear algebra operations, which makes it easy to
implement and parallelize.  Third, SCDM is a deterministic, one-step
procedure and does not require an initial guess. Hence, it will not get stuck at local minima as other nonconvex, nonlinear optimization
methods may. Finally, SCDM unifies the treatment of
molecules and crystals, while standard methods often require a
significantly more complex treatment for crystals.
We provide an interface to the widely used \texttt{Wannier90}
software package~\cite{MostofiYatesLeeEtAl2008} and demonstrate the effectiveness of SCDM via several examples of
real materials with isolated and entangled eigenvalues.

\section{The SCDM method}
\label{sec:scdm}
%\noindent\textbf{The SCDM method}

We consider the effective one-body Schr\"odinger operator $\mc{H} =
-\frac12 \Delta + V(\vr)$ in $\RR^{3}$, and all eigenfunctions of
interest $\psi_{i}(\vr)\in L^{2}(\RR^3)$. This corresponds to problems involving molecules and nanoclusters, which require a simpler setup than our forthcoming discussion of crystals.

For the isolated case, without
loss of generality we assume only the algebraically smallest $N$ eigenvalues
$\{\varepsilon_{i}\}_{i=1}^{N}$ are in the interval $\mc{I}$, and the
corresponding eigenfunctions $\{\psi_{i}\}_{i=1}^{N}$ are
orthonormal.
Using Dirac notation, the \emph{density matrix}  
\[
P =
\sum_{i=1}^{N} \ket{\psi_{i}}\bra{\psi_{i}}
\] is a rank $N$ matrix that is
the spectral projector associated with $\mc{H}$ onto the interval $\mc{I}.$ Notably, its
kernel $P(\vr,\vr')$ decays rapidly as
$\abs{\vr-\vr'}\to\infty$ \response{(specifics of the decay rates may be found in,
\eg,~\cite{BenziBoitoRazouk2013,Kohn1996})}.
Intuitively, if we can select a set of 
$N$ points $\mc{C}=\{\vr_{i}\}_{i=1}^{N}$ so that the corresponding
column vectors of the kernel $\{P(\vr,\vr_{i})\}_{i=1}^{N}$ are the ``most
representative'' and well conditioned column vectors of $P$, these vectors almost form the desired Wannier functions up to the orthonormality condition.

In order to select the set $\mc{C}$, we let $\Psi\in\CC^{N_{g}\times
N}$ denote the unitary matrix corresponding to a discrete representation of 
$\{\psi_{i}(\vr)\}_{i=1}^{N}$ using their nodal values on
$N_{g}$ grid points\footnote{\response{We are implicitly assuming $N_g \geq N,$ and since we have discretized $\vr$ the set $\mc{C}$ will correspond to picking $N$ of the $N_g$ points to define the columns.}}. The corresponding discretized density matrix, still denoted by
$P$, is given by $P=\Psi \Psi^{*}$. 
Conceptually, the most representative column
vectors can be identified via a QR factorization with column-pivoting
(QRCP) \cite{GolubVan2013} applied to
$P$. However, this is often impractical since
$P$ is prohibitively expensive even to construct and store in memory. 
The SCDM method~\cite{DamleLinYing2015} \response{leverages the fact that a good} set
$\mc{C}$ can be equivalently computed via
the QRCP of the matrix $\Psi^{*}$ \response{(see Remark~\ref{rem:qrcp})} as 
\begin{equation}
  \Psi^{*} \Pi = QR \equiv Q \begin{bmatrix} R_1 & R_2\end{bmatrix}.
  \label{eqn:qrcp}
\end{equation}
Here $\Pi$ is a permutation matrix, $Q$ is a unitary matrix,
$R_1\in \mathbb{C}^{N\times N}$ is an upper triangular matrix, and
$R_2\in\mathbb{C}^{N\times(N_{g}-N)}$. The points
$\mc{C}=\{\vr_{i}\}_{i=1}^{N}$ can be directly identified from the first
$N$ columns of the permutation matrix $\Pi.$

\response{
\begin{remark}
\label{rem:qrcp}
Under the assumption that the QRCP is computed via the algorithm by
Golub and Businger~\cite{businger1965linear}, in exact arithmetic the
permutation matrices $\Pi$ computed for factorizations of $P$ and
$\Psi^{*}$ will be identical. However, there exist other algorithms for
computing so-called rank-revealing QR
factorizations~\cite{GuEisenstat1996,chandrasekaran1994rank}. Therefore, the more relevant aspect of SCDM is that if a good rank-revealing QR factorization $\Psi^{*} \Pi = Q \begin{bmatrix} R_1 & R_2\end{bmatrix}$ is computed\textemdash in that $R_1$ is well conditioned\textemdash we have a good rank-revealing QR factorization of $P$ as $P \Pi = \left(\Psi Q\right) \begin{bmatrix} R_1 & R_2\end{bmatrix}.$ This justifies our use of $\Psi^{*}$ independent of the actual algorithm used for the rank-revealing QR factorization.
\end{remark}
}

Having chosen $\mc{C}$, we must now orthonormalize the localized column vectors
$\{P(\vr,\vr_{i})\}_{i=1}^{N}$ without destroying their locality. Note that 
\[
P(\vr,\vr_{i})=\sum_{i'=1}^{N} \psi_{i'}(\vr)\Xi_{i',i}
\]
where $\Xi\in\CC^{N\times N}$ h{}as matrix elements
$\Xi_{i,i'} = \psi^{*}_{i}(\vr_{i'})$.
One way to enforce the orthogonality is to define
\begin{equation}
  w_{i}(\vr) = \sum_{i'=1}^{N} \psi_{i'}(\vr) U_{i',i}, \quad U = \Xi
  (\Xi^{*}\Xi)^{-\frac12}.
  \label{eqn:Ugauge}
\end{equation}
Here $U\in \CC^{N\times N}$ is a unitary matrix and is referred to as a
\emph{gauge} in the physics literature. The matrix square root
transformation in Eq.~\eqref{eqn:Ugauge} is called the L\"owdin
transformation~\cite{Loewdin1950} and may be equivalently computed using the orthogonal factors from the reduced SVD of $\Xi$.

Considering
\begin{equation}
  (\Xi^{*}\Xi)_{i,i'} = \sum_{i''=1}^{N} \psi_{i''}(\vr_{i})
  \psi^*_{i''}(\vr_{i'}) = P(\vr_{i},\vr_{i'}),
  \label{eqn:XiXi}
\end{equation}
the decay properties of $P$ imply that $[P(\vr_{i},\vr_{i'})]$
may be viewed as a localized $N\times N$ matrix. If the eigenvalues of
$\left(\Xi^{*}\Xi\right)^{-\frac12}$ are bounded from below by a
positive value, then $\left(\Xi^{*}\Xi\right)^{-\frac12}$ will itself be localized~\cite{BenziBoitoRazouk2013}, and consequently
$\{w_{i}\}_{i=1}^{N}$ will be
localized, orthonormal Wannier functions.

\response{
\begin{remark}
Numerical observations indicate that for many real materials, the eigenvalues of $\left(\Xi^{*}\Xi\right)^{-\frac12}$
are indeed bounded from below by a positive value. Furthermore, the
condition number of this matrix can be very close to $1$ in practice. However, 
it is known that topological insulators (see,
\emph{e.g.},~\cite{Haldane1988Hall,chang2015high}) have isolated band
structure but do not admit exponentially localized
Wannier functions. Hence, there must necessarily be a failure mode of our
algorithm. In fact, for the topologically nontrivial Kane-Mele
model~\cite{Kane2005KaneMele} numerical experiments have shown that,
when restricted to the occupied bands, irrespective of the column set
used $\Xi^{*}(\vk)\Xi(\vk)$ in Eq.~\eqref{eqn:Uscdm} will
become singular for some $\vk$ in the Brillouin
zone~\cite{Soluyanov2011WannierZ2}.  Therefore, the SCDM method,
predicated on being able to choose columns good for all $\vk,$
necessarily fails. Such a statement holds generally for topological
insulators, such as Chern insulators and $\mathbb{Z}_{2}$
insulators~\cite{BrouderPanatiCalandraEtAl2007,Cornean2017WannierTRS}.
\end{remark}
}

For the entangled case, 
we extend the SCDM method by ``entangling'' the
eigenfunctions of interest with additional eigenfunctions
through the use of a quasi-density matrix
\begin{equation}
  P = \sum_{i} \ket{\psi_{i}}f(\varepsilon_{i})\bra{\psi_{i}} = f(H),
  \label{eqn:quasi_density}
\end{equation}
where $f(\cdot)$ is a smooth function, $\mc{I}$ is a subset of the
support set of $f$, and the summation is formally over \emph{all}
eigenfunctions of $\mc{H}$.
From this perspective, the case of isolated band is associated with the choice
%\begin{equation}
$f(\varepsilon) = \mathbf{1}_{\mc{I}}(\varepsilon)$,
%  \label{eqn:fisolate}
%\end{equation}
%where $\mathbf{1}_{\mc{I}}$ is 
the indicator function on the interval $\mc{I}$. 

\begin{figure}[ht]
  \centering 
  \includegraphics[width=.8\columnwidth]{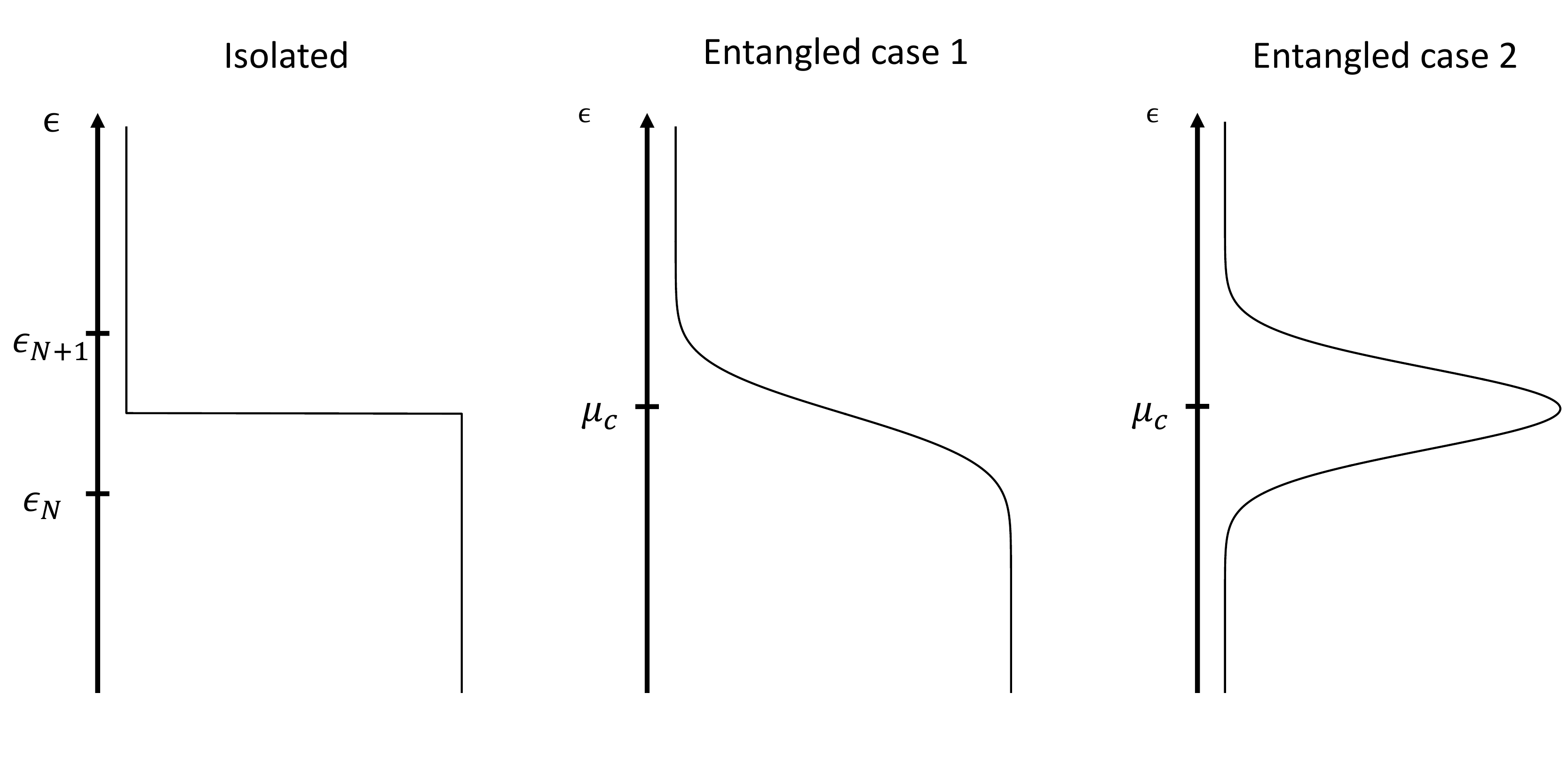}
  \caption{$f(\varepsilon)$ for the isolated and two entangled cases.}
  \label{fig:fepsilon}
\end{figure}

We now assume there is a number $\mu_c$ such that 
$\inf_{i} \abs{\varepsilon_{i}-\mu_{c}}$ is very small or even zero. 
The following two scenarios of entangled eigenvalues appear most
frequently in quantum physics, corresponding to
the Wannier localization problem below and around a certain energy level
(usually the Fermi energy)
respectively~\cite{YatesWangVanderbiltEtAl2007}. In both cases $f(\varepsilon)$ is large on the
region of interest and smoothly decays to zero outside $\mathcal{I}$ in a manner controlled by a parameter $\sigma$
(see Fig.~\ref{fig:fepsilon}). 

\noindent\textbf{Entangled case 1:} $\mc{I}=(-\infty,\mu_{c})$.
%This corresponds to the Wannier localization problem for metallic
%systems as well as insulating systems with conduction bands.
In this case we can choose a value $\sigma>0$ and let
\begin{equation}
  f(\varepsilon) = \frac12 \mathrm{erfc}\left(
  \frac{\varepsilon-\mu_{c}}{\sigma} \right)=\frac{1}{\sqrt{\pi\sigma^2}}\int_{\varepsilon}^{\infty} \exp\left(-\frac{(t-\mu_c)^2}{\sigma^2}\right)\ud t.
  \label{eqn:ferfc}
\end{equation}
The function $f(\varepsilon)$ satisfies
$\lim_{\varepsilon\to-\infty}f(\varepsilon)=1$,  
$\lim_{\varepsilon\to\infty}f(\varepsilon) = 0$ and the transition
occurs smoothly around $\mu_{c}$.  

\noindent\textbf{Entangled case 2:} $\mc{I}=(\mu_{c}-\sigma,\mu_{c}+\sigma)$.  
%This corresponds to the Wannier localization problem near the Fermi energy.
In this case we choose $f$ to be a
Gaussian function
\begin{equation}
  f(\varepsilon) = \exp\left(
  -\frac{(\varepsilon-\mu_{c})^2}{\sigma^2}
  \right).
  \label{eqn:fgaussian}
\end{equation}

%\response{While in both cases $f(\varepsilon)$ exhibits the properties we need\textemdash smoothness and decay\textemdash they are not necessarily optimal under some formal criteria, and therefore not the only possible choices. Nevertheless, as we will demonstrate they work remarkably well in practice.}

\response{In both cases $f(\varepsilon)$ exhibits the smoothness and
decay properties that we need, but the choice of $f(\varepsilon)$ is
certainly not unique. For instance, in the entangled case 1 we may use
the Fermi-Dirac function instead. Nevertheless, numerical results
indicate that the present choices perform well in practice.}

\begin{figure}[ht]
\centering
    \includegraphics[width=.8\columnwidth]{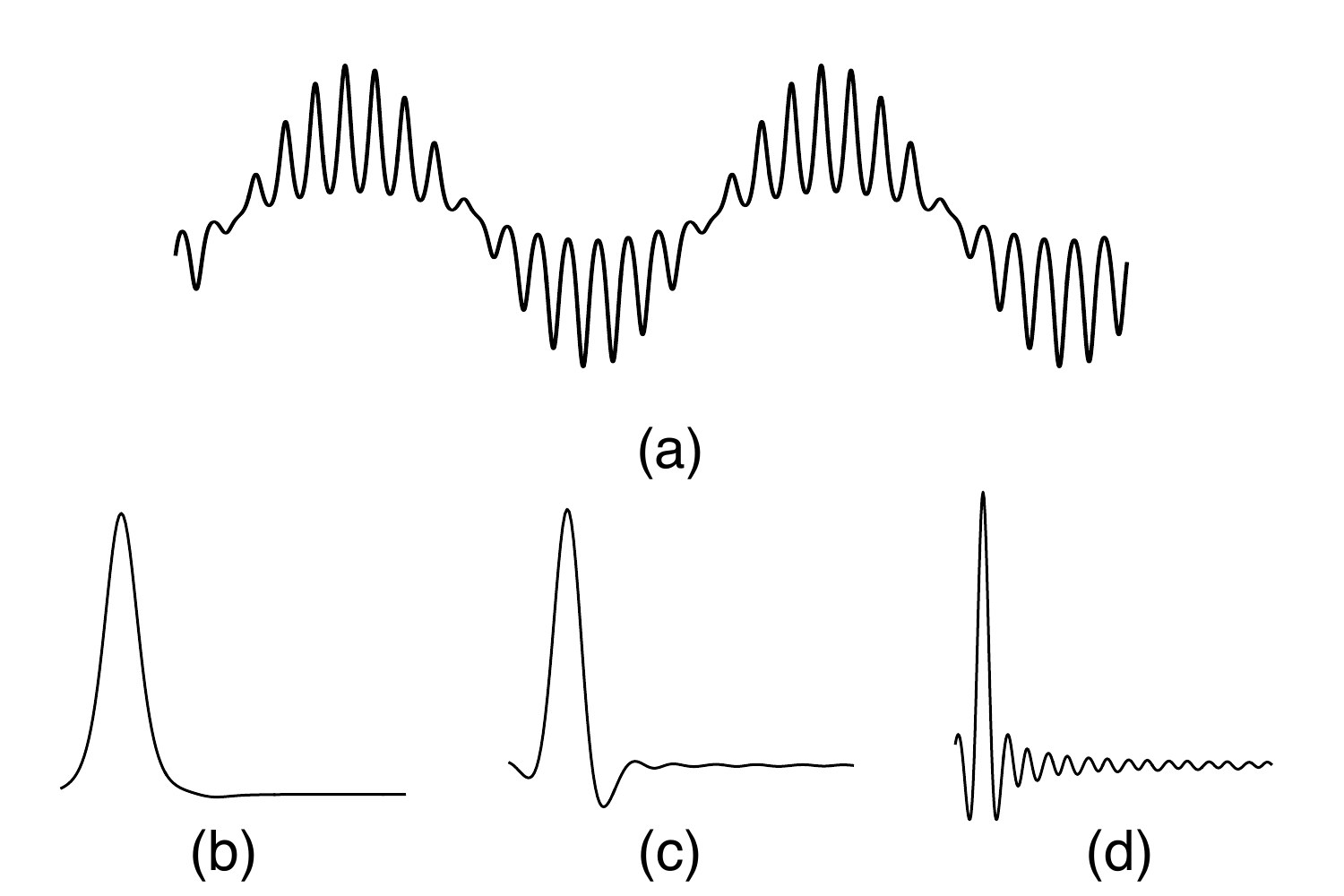}
  \caption{Eigenfunctions and computed Wannier functions
  for a simple one-dimensional model problem. (a) An example eigenfunction plotted on the whole domain,
  (b) the isolated case, (c) entangled case 1, and (d) entangled case 2. For the three examples of functions computed by the SCDM method we have zoomed in on the region where the bulk of the function is supported.}
  \label{fig:model_example}
\end{figure}

For a smooth function $f$, the kernel of the quasi-density matrix
$P(\vr,\vr')$ also decays rapidly\footnote{\response{Here we deliberately omit
the discussion on the decay rate in order to unify the discussion for
molecular and crystal systems. For the molecular case here, we may apply the theoretical
statements in \eg~\cite{BenziBoitoRazouk2013,Lin2017}}}. \response{Given a desired number of Wannier functions $N_w$\footnote{\response{For metallic systems this is often set to be equal to the number of bands plus a small integer, a heuristic recently justified mathematically~\cite{Cornean2017wanniermetalic}}}} we would once again like to select $N_w$ ``most representative'' and well conditioned column vectors of $P$ to construct them. Let  $\mc{E}=\mathrm{diag}\left[ \left\{ \varepsilon_{i}
\right\}\right]\in \RR^{N\times N}$ be a diagonal matrix  containing all eigenvalues such that $f(\varepsilon)$ is above some threshold, and $\Psi\in\CC^{N_g\times N}$ be the matrix containing the corresponding discretized eigenvectors. We can now compute a QRCP for the weighted eigenvectors 
\begin{equation}
  \left( f(\mc{E})\Psi^{*} \right)\Pi = QR
  \label{eqn:qrcp_weight}
\end{equation}
and select the $N_{w}$ columns corresponding to the left most $N_{w}$ columns
of the permutation $\Pi$. As before, we let $\mc{C}=\{\vr_{i}\}_{i=1}^{N_{w}}$ denote the real space points corresponding
to the selected columns and define the auxiliary matrix
$\Xi\in\CC^{N\times N_{w}}$ with $\Xi_{i,i'} =
f(\varepsilon_{i})\psi^{*}_{i}(\vr_{i'})$.
If the eigenvalues
of $\Xi^{*}\Xi$ are bounded away from $0$, the choice of gauge
$U=\Xi (\Xi^{*}\Xi)^{-\frac12}$  once again gives rise to the Wannier functions. Now,
$U\in \CC^{N\times N_{w}}$ is a rectangular matrix with orthonormal columns.
Fig.~\ref{fig:model_example} compares the delocalized eigenfunctions
and the localized Wannier functions corresponding
to isolated and entangled cases using a simple one-dimensional model
problem, \response{the details of which may be found in Appendix~\ref{a:onedim}}.

%Using these ideas, the SCDM procedure handles both the isolated and the entangled
%case.  Instead of ``disentanglement'', our method ``entangles'' the
%eigenfunctions of interest with additional eigenfunctions
%through a quasi-density matrix and uses it to directly construct the Wannier functions. 

%Figure~\ref{fig:model_example} shows examples for Wannier functions constructed, in each of the three cases, for a simple one-dimensional model problem with a periodic well potential. Here, only a portion of the computational domain is shown to more clearly illustrate the local structure of the Wannier functions. The axis discrepancy is a result of the locality\textemdash both the eigenfunctions and Wannier functions have norm one, so given the locality the Wannier functions are larger in magnitude on their support. 

% %%%%%%%%%%%%%%%%%%%%%%%%%%%%%%%%%%%%%%%%%%%%%%%%%%%%%%%%%%%%%%%%%%%%%%%%%%%%%%%
\section{Bloch-Floquet theory}\label{subsec:brillouin}
%Algorithm for finding Wannier functions becomes significantly more involved
%when translational invariance of the crystal structure is taken into
%account.  From this perspective, the prior
%example can be considered as a special case of crystalline structure.
To facilitate further discussion we briefly review
Bloch-Floquet theory for crystal structures. Without loss of generality
we consider a three-dimensional crystal. 
The \emph{Bravais lattice} with lattice vectors $\va_{1},\va_{2},\va_{3}\in \RR^{3}$ is defined as
\begin{equation}
  \mathbb{L} = \left\{ \vR \vert \vR = n_1\bvec{a}_1 + n_2 \bvec{a}_2 +
n_3\bvec{a}_3, \quad n_1,n_2,n_3\in\mathbb{Z}\right\}.
  \label{eqn:Blattice}
\end{equation}
The the potential $V$ is real-valued and $\mathbb{L}$-periodic, \ie~ 
\[
V(\vr+n_{i}\va_{i}) = V(\vr), \quad \forall \vr\in \RR^3, n_{i}\in
\ZZ.
\]
%\begin{equation}
%  V(\vr+n_{i}\va_{i}) = V(\vr), \quad \forall \vr\in \RR^3, n_{i}\in
%  \ZZ, \quad
%  i=1,2,3.
%  \label{eqn:Vperiod}
%\end{equation} 
The unit cell is defined as
\begin{equation}
 \Gamma =\left\{ \vr = c_{1} \va_{1} + c_{2} \va_{2} + c_{3}
 \va_{3} \vert -1/2 \le c_{1},c_{2},c_{3}< 1/2\right\}.
 \label{eqn:unit_cell}
\end{equation}
The Bravis lattice induces a reciprocal lattice
$\mathbb{L}^{*}$, and the unit cell of the reciprocal lattice is
called the (first) \emph{Brillouin zone} and denoted by
$\Gamma^{*}$. The problem formulation in the previous section can be formally identified as a special
case of this setup with an infinitely large unit cell.
%The reciprocal lattice vectors $\bvec{b}_1,\bvec{b}_2,\bvec{b}_3$
%satisfy $\bvec{a}_{i} \cdot \bvec{b}_{j} = 2\pi
%\delta_{ij}, i,j=1,2,3$, which induces a reciprocal lattice
%$\mathbb{L}^{*}$. The unit cell of the reciprocal lattice is
%called the (first) \emph{Brillouin zone}, defined as
%\[
%\Gamma^{*} = \left\{ \vk = k_{1} \vb_{1} + k_{2} \vb_{2} + k_{3}
%  \vb_{3} \Big\vert -\frac12 \le k_{1},k_{2},k_{3}< \frac12\right\}.
%\]
%The Brillouin zone has a number of special points related to the
%symmetry of the crystal. The common special point is the
%$\Gamma$-point\footnote{notation in crystallography, and not to be confused with $\Gamma$ used to denote
%the unit cell in this paper}, which corresponds to $\vk=(0,0,0)^{T}$.

According to the Bloch-Floquet theory, the spectrum of $\mc{H}$ can be
relabeled using two indices $(b,\vk)$, where $b\in \mathbb{N}$ is called the band
index and $\vk\in\Gamma^{*}$ is the Brillouin zone index.
Each generalized eigenfunction $\psi_{b,\vk}(\vr)$ is known as a Bloch orbital and 
satisfies
$\mc{H}\psi_{b,\vk}(\vr)=\varepsilon_{b,\vk}\psi_{b,\vk}(\vr)$ \response{with periodic boundary conditions.} Furthermore,
$\psi_{b,\vk}$ can be decomposed as $\psi_{b,\vk}(\vr) = e^{\I \vk\cdot
\vr} u_{b,\vk}(\vr)$,
where $u_{b,\vk}(\vr)$ is a periodic function with respect to
$\mathbb{L}$. An eigenpair $(\varepsilon_{b,\vk},u_{b,\vk})$ can be 
%the periodic part of $\psi_{b,\vk}(\vr)$ satisfying
%\begin{equation}
%  u_{b,\vk}(\vr+\vR) = u_{b,\vk}(\vr), \quad \forall \vR\in\Gamma,
%  \label{eqn:upbc}
%\end{equation}
obtained by solving the eigenvalue problem
\begin{equation}
  \mc{H}(\vk) u_{b,\vk} = \varepsilon_{b,\vk} u_{b,\vk}(\vr), \quad \vr\in \Gamma, \quad \vk \in
  \Gamma^{*},
  \label{eqn:bandproblem}
\end{equation}
where $\mc{H}(\vk) = -\frac12 (\nabla+\I \vk)^2 + V(\vr)$.
For each $\vk$, the eigenvalues $\varepsilon_{b,\vk}$ are ordered
non-decreasingly. For a fixed $b$, $\{\varepsilon_{b,\vk}\}$ as a
function of $\vk$ is called a
\response{\emph{Bloch band}}. The collection of all eigenvalues are called the
\emph{band structure} of the crystal, which characterizes the spectrum
of the operator $\mc{H}$. In this framework, the isolation condition~\eqref{eqn:isolation}
becomes
\begin{equation}
%  \inf_{\stackrel{\vk,\vk'\in\Gamma^*}{\varepsilon_{b,\vk}\in
%  \mc{I},\varepsilon_{b',\vk'}\notin \mc{I}}}
%  \abs{\varepsilon_{b,\vk}-\varepsilon_{b',\vk'}}>0.
  \inf
  \abs{\varepsilon_{b,\vk}-\varepsilon_{b',\vk'}}>0, \quad
%  \text{s.t.}~
  \vk,\vk'\in\Gamma^*, \varepsilon_{b,\vk}\in \mc{I},\varepsilon_{b',\vk'}\notin \mc{I}.
  \label{eqn:gapcondition}
\end{equation}

\section{Wannier functions for crystals and disentanglement}
Mirroring our prior discussion, we first we consider the isolated case. Without loss of generality we assume the eigenvalues in $\mc{I}$ are labeled as 
$\{\varepsilon_{b,\vk}\}_{b=1}^{N_{b}}$.
If we rotate $\{\psi_{b,\vk}\}$ by an arbitrary
unitary matrix, now indexed by $\vk,$ $U(\vk)\in\CC^{N_{b}\times N_{b}}$, we can
define a new set of functions
\begin{equation}
  \wt{\psi}_{b,\vk}(\vr) = \sum_{b'=1}^{N_{b}} \psi_{b',\vk}(\vr) U_{b',b}(\vk), \quad
  \vk\in \Gamma^{*}.
  \label{eqn:Utransform}
\end{equation}
A given set of matrices $\{U(\vk)\}$ is called a \emph{Bloch gauge}.
For any choice of Block gauge, the Wannier
functions for crystals are~\cite{Wannier1937}
\begin{equation}
  w_{b,\vR}(\vr) = \frac{1}{|\Gamma^{*}|}\int_{\Gamma^{*}} \wt{\psi}_{b,\vk}(\vr)
  e^{-\I \vk\cdot \vR}
  \ud \vk, \quad \vr\in\RR^3, \vR\in \mathbb{L},
%  \varepsilon_{b,\vk} \in \mc{I}, 
  \label{eqn:wannier}
\end{equation}
where $|\Gamma^{*}|$ is the volume of the Brillouin zone.
%When the isolation condition \eqref{eqn:gapcondition} is satisfied, 
For a class of $\mc{H}$ there
exists a gauge such that $\wt{\psi}_{b,\vk}$ is analytic in
$\vk$, implying that each Wannier function decays exponentially as
$\abs{\vr}\to \infty$~\cite{Blount1962,PanatiPisante2013}. 
Furthermore, the set of Wannier functions 
$\{w_{b,\vR}(\vr)\}$ forms an
orthonormal basis of the subspace in $L^{2}(\RR^3)$ spanned by the Bloch
orbitals associated with eigenvalues in $\mc{I}$. 
%More specifically,
%\begin{equation}
%  \wt{\psi}_{b,\vk}(\vr) = \sum_{\vR\in \mathbb{L}} e^{\I \vk\cdot\vR}
%  w_{b,\vR}(\vr).
%  \label{eqn:blochwannier}
%\end{equation}
For crystals, the Wannier localization problem is thus partly
reduced to the problem of finding a gauge such that $\wt{\psi}_{b,\vk}$ is
smooth with respect to $\vk$. This can be done by minimizing the
``spread functional''~\cite{MarzariVanderbilt1997}
\begin{equation}
  \Omega[\{w_{b,\bvec{0}}\}_{b=1}^{N_{b}}] = \sum_{b=1}^{N} 
   \int \abs{w_{b,\bvec{0}}(\vr)^2} \vr^2 \ud \vr - \left( 
   \int \abs{w_{b,\bvec{0}}(\vr)^2} \vr \ud \vr \right)^2.
   \label{eqn:spread}
\end{equation}
Only $\vR=\bvec{0}$ is considered because Wannier functions associated with
different $\vR$'s only differ by translation.

In the entangled case, the disentanglement
method constructs the gauge via a two-step procedure. It
first finds a gauge $U^{\text{dis}}(\vk)\in \CC^{N_{b}\times N_{w}}$, in order to disentangle the given set of $N_b$ functions into $N_w$ functions for each $\vk$.  
This is obtained by \response{minimizing} a ``smoothness
functional'' \response{(the gauge invariant part of the spread
functional of $\Omega$,
see~\cite{SouzaMarzariVanderbilt2001,MarzariMostofiYatesEtAl2012} for
details)} with respect to $\vk$ for
the following auxiliary  functions
\begin{equation}
  \psi^{\text{dis}}_{b,\vk}(\vr) = \sum_{b'=1}^{N_{b}} \psi_{b',\vk}(\vr)
  U^{\text{dis}}_{b',b}(\vk), \quad b=1,\ldots,N_{w},
  \vk\in \Gamma^{*}.
  \label{eqn:psidis}
\end{equation}
After obtaining $U^{\text{dis}}(\vk)$
a second gauge $U^{\text{loc}}(\vk)\in \CC^{N_{w}\times N_{w}}$ for
each $\vk$ is computed by minimizing the spread
functional~\eqref{eqn:spread}. Finally, the overall gauge in the
disentanglement method is the composition $U(\vk) =  U^{\text{dis}}(\vk)
U^{\text{loc}}(\vk)$.
%\begin{equation}
%  
%  \label{eqn:wanniergauge}
%\end{equation}

This two-step procedure can be viewed as a heuristic means to
solve a nonlinear, nonconvex multi-objective optimization problem aiming to simultaneously
maximize the smoothness functional with respect to $\vk$, and minimize the spread
functional. Our numerical results indicate that, at least
in some cases, this two step procedure may not be an effective surrogate
for the desired optimization problem.

\section{SCDM for crystals}

We now proceed to discuss the relatively minor modifications needed to generalize the SCDM method to crystals. 

In the isolated case, for each $\vk$-point in the Brillouin zone the
$\vk$-dependent density matrix is gauge invariant
\begin{equation}
  P(\vk) = \sum_{\varepsilon_{b,\vk}\in \mc{I}}
  \ket{\psi_{b,\vk}}\bra{\psi_{b,\vk}} = \sum_{\varepsilon_{b,\vk}\in \mc{I}}
  \ket{\wt{\psi}_{b,\vk}}\bra{\wt{\psi}_{b,\vk}},
  \label{eqn:projector}
\end{equation}
and is already an
analytic function of $\vk$~\cite{Nenciu1991,PanatiPisante2013}. The
SCDM method uses the density matrix to construct a gauge so that
$\wt{\psi}_{b,\vk}$ is smooth with respect to $\vk$.
However, for crystals, we need to select a common set of columns for all the
$\vk$-dependent density matrices.  Previous
work~\cite{DamleLinYing2017} suggests that it is often sufficient to select the
columns using an ``anchor'' point $\vk_{0}$, such as the so-called Gamma-point
$(0,0,0)^{T}$ to identify these columns. A generalization of this procedure is also outlined in \cite{DamleLinYing2017}, though in our numerical experiments we have found use of the Gamma-point as the anchor point suffices.

Let $\Psi_{\vk}\in\CC^{N_{g}\times
N_{b}}$ be the unitary matrix representing $\{\psi_{b,\vk}(\vr)\}$ on a discrete grid
in the unit cell. At the anchor point $\vk_{0}$, we compute the QRCP
\[
\Psi_{\vk_{0}}^{*} \Pi = Q R.
\]
As before, let $\mc{C}=\{\vr_{b}\}_{b=1}^{N_{b}}$ denote the grid points corresponding to the $N_{b}$ selected columns where
$\vr_{b}\in\Gamma$. 
For each $\vk$ point, define the auxiliary matrix $\Xi\in\CC^{N_{b}\times
N_{b}}$ with matrix elements $\Xi_{b,b'}(\vk) =
\psi^{*}_{b,\vk}(\vr_{b'})$.
Then the smoothness of the density matrix $P(\vk)$
implies that each function 
\[
P_{b,\vk}(\vr) = \sum_{b'=1}^{N_{b}} \psi_{b',\vk}(\vr)
\Xi_{b',b}(\vk)
\]
is smooth with respect to $\vk$.  
%In order to construct
%$\{\wt{\psi}_{b,\vk}\}$ via a unitary transformation of
%$\{\wt{\psi}_{b,\vk}\}$, we can 
%What remains is to orthogonalize these functions and construct a unitary gauge.
As before the SCDM gauge can be constructed via the L\"owdin
transformation as
\begin{equation}
  U(\vk) = \Xi(\vk) \left[\Xi^{*}(\vk)\Xi(\vk)\right]^{-\frac12}.
  \label{eqn:Uscdm}
\end{equation}
Similar to Eq.~\eqref{eqn:XiXi},
\begin{equation}
  (\Xi^{*}(\vk)\Xi(\vk))_{b,b'} = \sum_{b''=1}^{N_{b}} \psi_{b'',\vk}(\vr_{b})
  \psi^*_{b'',\vk}(\vr_{b'}) = P(\vr_{b},\vr_{b'};\vk).
\end{equation}
Since the kernel $P(\vr_{b},\vr_{b'};\vk)$ is smooth with respect to
$\vk$, \response{a sufficient condition for the matrix
$\left(\Xi^{*}(\vk)\Xi(\vk)\right)^{-\frac12}$ to be smooth
with respect to $\vk$ is that 
$\|\Xi^{*}(\vk)\Xi(\vk)-I\|_{2} \leq \frac{1}{2}$ for all $\vk$ in the Brillouin zone.  As discussed
before, numerical observations for many real materials indicate that
this condition holds, but would fail in the context of topological
insulators.}

\response{
\begin{remark}
Although the SCDM algorithm for the isolated case would fail for
topological insulators, when the Hamiltonian satisfies the time-reversal
symmetry, it is still of interest to modify the SCDM method to select
``generalized columns'' to construct well localized Wannier
functions that do not obey the time-reversal
symmetry~\cite{Cornean2017WannierTRS}.  Another interesting possibility
for treating general topological insulators is to increase the number of
Wannier functions and use the algorithm for the entangled case. This
will be investigated in the future.
\end{remark}
}

As a result,
\begin{equation}
  \wt{\psi}_{b,\vk}(\vr) = \sum_{b'=1}^{N_{b}} \psi_{b',\vk}(\vr) U_{b',b}(\vk) = 
  \sum_{b'=1}^{N_{b}} P_{b',\vk}(\vr)
  \left[\Xi^{*}(\vk)\Xi(\vk)\right]^{-\frac12}_{b',b}
  \label{eqn:psismoothk}
\end{equation}
is smooth with respect to $\vk$.  Therefore, the Fourier transform of
$\wt{\psi}_{b,\vk}$  with respect to $\vk$ as in Eq.~\eqref{eqn:wannier} gives rise to the Wannier
functions for crystals constructed by the SCDM method.

In the entangled case (when \eqref{eqn:gapcondition} is not satisfied), the SCDM method makes use of a $\vk$-dependent quasi-density matrix for each $\vk$ point:
\begin{equation}
  P(\vk) = \sum_{\varepsilon_{b,\vk}} 
  \ket{\psi_{b,\vk}}f(\varepsilon_{b,\vk})\bra{\psi_{b,\vk}}. \label{eqn:projectorentangle}
\end{equation}
Here, the choice of $f(\varepsilon)$ matches what we previously discussed, and depends on the desired $\mc{I}.$
In particular, the reduction to the isolated case is again the choice $f(\varepsilon) =
\mathbf{1}_{\mc{I}}(\varepsilon)$, which reduces the $\vk$-dependent quasi-density matrix to a $\vk$-dependent density matrix.
Numerical results indicate that these quasi-density matrices for the two
entangled cases are smooth with respect to $\vk$.%, although we are not aware of theoretical results in the literature.

Let $\mc{E}(\vk)=\mathrm{diag}\left[ \left\{ \varepsilon_{b,\vk}
\right\}_{b=1}^{N_{b}} \right]$ be a diagonal matrix for each $\vk$ containing eigenvalues such that $f(\varepsilon)$ is larger than some threshold.
Computing a QRCP at the anchor point in the Brillouin zone $\vk_{0}$, we obtain
\begin{equation}
  f(\mc{E}(\vk_{0}))\Psi_{\vk_{0}}^{*}\Pi = QR.
  \label{eqn:qrcp_anchor}
\end{equation}
Analogously to before, the set of real space points $\mc{C}=\{\vr_{b}\}_{b=1}^{N_{w}}$ 
are given by the left most $N_{w}$ columns
of the permutation matrix $\Pi$. Defining the auxiliary matrix
$\Xi(\vk)\in\CC^{N_{b}\times N_{w}}$ with matrix elements 
\[
\Xi_{b,b'}(\vk) = f(\varepsilon_{b,\vk})\psi^{*}_{b,\vk}(\vr_{b'})
\]
implies that
\[
 P_{b,\vk}(\vr) = \sum_{b'} \psi_{b',\vk}(\vr) \Xi_{b',b}(\vk)
\]
is smooth with respect to $\vk$. If the eigenvalues 
of $[\Xi^{*}(\vk)\Xi(\vk)]$ are uniformly bounded away from $0$ in the Brillouin zone, the
gauge $U(\vk)\in \CC^{N_{b}\times N_{w}}$ given by Eq.~\eqref{eqn:Uscdm}
is unitary and via \eqref{eqn:psismoothk} defines
$\{\wt{\psi}_{b,\vk}\}$ that are smooth with respect to $\vk$. 
Eq.~\eqref{eqn:wannier} once again yields the desired Wannier functions.

%%%%%%%%%%%%%%%%%%%%%%%%%%%%%%%%%%%%%%%%%%%%%%%%%%
\section{Wannier interpolation for band structure}
%%%%%%%%%%%%%%%%%%%%%%%%%%%%%%%%%%%%%%%%%%%%%%%%%%

In practical electronic structure calculations, the Brillouin zone needs
to be discretized using a finite number of points denoted by the set
$\mc{K}$. The most widely used
discretization scheme is the Monkhorst-Pack
grid~\cite{MonkhorstPack1976}, which corresponds to a uniform discretization of
$\Gamma^{*}$. However,
the band structure $\varepsilon_{b,\vk}$ as a
function of $\vk$ often needs to be computed on finely discretized paths (not necessarily grid aligned) through the Brillouin zone. Because the
eigenvalues $\varepsilon_{b,\vk}$ are in general only Lipschitz continuous with respect to
$\vk$~\cite{ReedSimonIV1978},
interpolating $\varepsilon_{b,\vk}$ for $\vk \notin \mc{K}$ directly
from the eigenvalues computed on the grid $\mc{K}$ can result in large
interpolation errors. The Wannier interpolation
method (see \eg,~\cite{MarzariMostofiYatesEtAl2012}), makes use of the locality of Wannier functions and can
yield both higher quality interpolation for a fixed $\mc{K}$ and improved convergence with respect to the
number of discretization points in $\mc{K}$. We simply pair our localized functions computed via the SCDM method with standard Wannier interpolation techniques to compute band structure.

Observe that the computed gauge (in this case from the SCDM method)
rotates the periodic part of the Bloch orbitals
as well according to 
\[
\wt{u}_{b,\vk}(\vr) = \sum_{b'} u_{b',\vk}(\vr)
U_{b',b}(\vk),
\]
and let $\wt{H}(\vk)$ denote the matrix representation of $\mc{H}(\vk)$ in the
basis $\{\wt{u}_{b,\vk}\}$. We may then construct the reduced matrix
\begin{equation}
  [\wt{H}(\vk)]_{b,b'} =
  \bra{\wt{u}_{b,\vk}}\mc{H}(\vk)\ket{\wt{u}_{b',\vk}} =
  [U^{*}(\vk)\mc{E}(\vk) U(\vk)]_{b,b'}.
  \label{eqn:Htconstruct}
\end{equation}
In particular, if the gauge were the identity matrix then $\wt{H}(\vk)$
would be a diagonal matrix with the eigenvalues at $\vk$ on the
diagonal. However, as noted earlier interpolation via this
representation may be very inaccurate. Rather, we would like to interpolate using matrices whose entries are smoother with respect to $\vk$ than the eigenvalues themselves are. 

This leads to the use of the Wannier functions in \eqref{eqn:Htconstruct}. The smoothness of $\wt{u}_{b,\vk}$ with respect to $\vk$ implies
that each entries of the matrix $\wt{H}$ is also smooth with respect to
$\vk$. Now, Wannier interpolation is precisely given by Fourier
interpolation of $\wt{H}(\vk)$ onto the desired points in $\Gamma^*.$ 
More specifically, the Fourier transform of $\wt{H}(\vk)$
\begin{equation}
  \mathfrak{H}(\vR) = \frac{1}{\abs{\Gamma^{*}}} \int 
  e^{-\I \vk\cdot \vR}\wt{H}(\vk)  \ud \vk
  \label{eqn:HR}
\end{equation}
decays rapidly as $\abs{\vR}\to\infty$. 
When the Monkhorst-Pack grid is used, $\mathfrak{H}(\vR)$ can be efficiently approximated as
\begin{equation}
  \mathfrak{H}(\vR) \approx \frac{1}{N_{\Gamma^{*}}} \sum_{\vk\in \mc{K}}e^{-\I \vk\cdot
  \vR} \wt{H}(\vk).
  \label{}
\end{equation}
Here $N_{\Gamma^{*}}=N_{1}N_{2}N_{3}$ is the total number of $\vk$-points.
From $\mathfrak{H}(\vR)$, 
$\wt{H}(\vk)$ for any $\vk\in\Gamma^{*}$ can be reconstructed as
\begin{equation}
  \wt{H}(\vk) = \sum_{\vR\in\mathbb{L}} e^{\I \vk \cdot \vR}
  \mathfrak{H}(\vR).
  \label{eqn:HK}
\end{equation}
This matrix is small with size $N_w \times N_w$, and computing its eigenvalues yields the interpolated band
structure $\varepsilon_{b,\vk}$ for $\vk\in\Gamma^*.$ 
The summation over the Bravis lattice in Eq.~\eqref{eqn:HK}
needs to be truncated.  The most natural
truncation of the Bravis lattice is the parallelepiped dual to the
Monkhorst-Pack grid. However, it has been \response{numerically} observed that the truncation with
a Wigner-Seitz cell leads to smaller numerical error especially when
$\mc{N}_{\Gamma^{*}}$ is relatively small~\cite{YatesWangVanderbiltEtAl2007}.

\section{Interface with \texttt{Wannier90}}

The solution of the Wannier localization problem is entirely encapsulated
in the Bloch gauge $\{U(\vk)\}.$ This allows us to easily integrate our SCDM method with the widely used
\texttt{Wannier90}~\cite{MostofiYatesLeeEtAl2008} software
package (available online at \url{www.wannier.org}) in a \textit{non-intrusive} way.
\texttt{Wannier90} requires an initial guess for the gauge, and we may simply provide ours as input using the proper file format. Similarly, we may use existing interfaces between electronic structure software packages and \texttt{Wannier90} to get the requisite input for our code. To facilitate our forthcoming numerical experiments, we
built an interface for our method to \texttt{Wannier90} and the code is available online\footnote{https://github.com/asdamle/SCDM}. 

This procedure allows us to leverage all the
functionalities of \texttt{Wannier90} directly. For example, we can then either set the number of iterations to zero, which allows us to both compute the spread of our SCDM based Wannier functions and perform Wannier interpolation using our computed gauge, or use our gauge as an initial guess and see if the optimization procedure is able to improve it. In addition, our strategy makes comparisons with Wannier functions computed by the optimization methodology in \texttt{Wannier90} simple.

%%%%%%%%%%%%%%%%%%%%%%%%%%%%%%%%%%%%%%%%%%%%%%%%%%%%%%%%%%%%%%%%%%%%%%%%%%%%%%%
\section{Numerical results}\label{sec:numer}
%%%%%%%%%%%%%%%%%%%%%%%%%%%%%%%%%%%%%%%%%%%%%%%%%%%%%%%%%%%%%%%%%%%%%%%%%%%%%%%

We now demonstrate the effectiveness of the SCDM method qualitatively and
quantitatively using real materials. The electronic structure
calculations are performed using the {\sc Quantum
ESPRESSO}~\cite{GiannozziBaroniBoniniEtAl2009} software
package with the PBE exchange-correlation
functionals~\cite{PerdewErnzerhofBurke1996}. 

Qualitatively, we examine the shape of the Wannier functions obtained
from SCDM, and compare against minimizer of the spread
functional~\eqref{eqn:spread} in \texttt{Wannier90}. Quantitatively, we
measure the value of the spread functional for Wannier functions
obtained from SCDM, as well as the accuracy of band structure
interpolation from the Wannier functions for isolated and entangled
cases. \response{For these examples, we consider the choice of $N_b,$
$N_w,$ and $N_g$ as fixed aspects of the problem instance and therefore
the SCDM methodology relies on at most two parameters\textemdash $\mu$
and $\sigma$\footnote{\response{While formally once could consider
the anchor point $\vk_0$ as a tunable parameter, we have observed
that the default choice of using the Gamma-point as the anchor point
performs robustly in practice. Similarly, the choice of $f(\varepsilon)$ is dictated by the type of problem that one wishes to solve.}}}.

% Practically, the Brillouin zone needs
% to be discretized using a finite number of points. However, $\varepsilon_{b,\vk}$ as a
% function of $\vk$ often needs to be computed on finely discretized paths
% through the Brillouin zone that do not coincide with the discretization
% grid\textemdash this is band structure (or Wannier) interpolation. 

% We
% built an interface for our method to the widely used \texttt{Wannier90}
% software package~\cite{MostofiYatesLeeEtAl2008} allowing us to use its interpolation routines. Our code is available online\footnote{https://github.com/asdamle/SCDM}.

% We first perform a simple test of our method by performing band interpolation for the four low-lying bands of Silicone. Here, the presence of a spectral gap means that we may use the most basic variant of our algorithm, the isolated case. Visually, in Figure~\ref{fig:si_low_kpt6} we observe very good agreement between the direct calculation and the Wannier interpolation via SCDM. The horizontal axis represents a standard path through $\Gamma^*$ on which the structure is desired. 

% \begin{figure}[ht]
%   \centering 
%   \includegraphics[width=0.5\columnwidth]{si_band_kpt6.eps}
%   \caption{(color online) Wannier interpolation with SCDM for the band structure of Silicone using a $6\times 6\times 6$ k-point grid. Direct calculation from QE (red line) and Wannier interpolation using SCDM (blue circles).}
%   \label{fig:si_low_kpt6}

% \end{figure}

% \todo{Demonstrate the invertibility of $\Xi(\vk)^* \Xi(\vk)$.}

\begin{figure}[ht]
\centering
    \subfloat[]{\includegraphics[width=0.55\columnwidth]{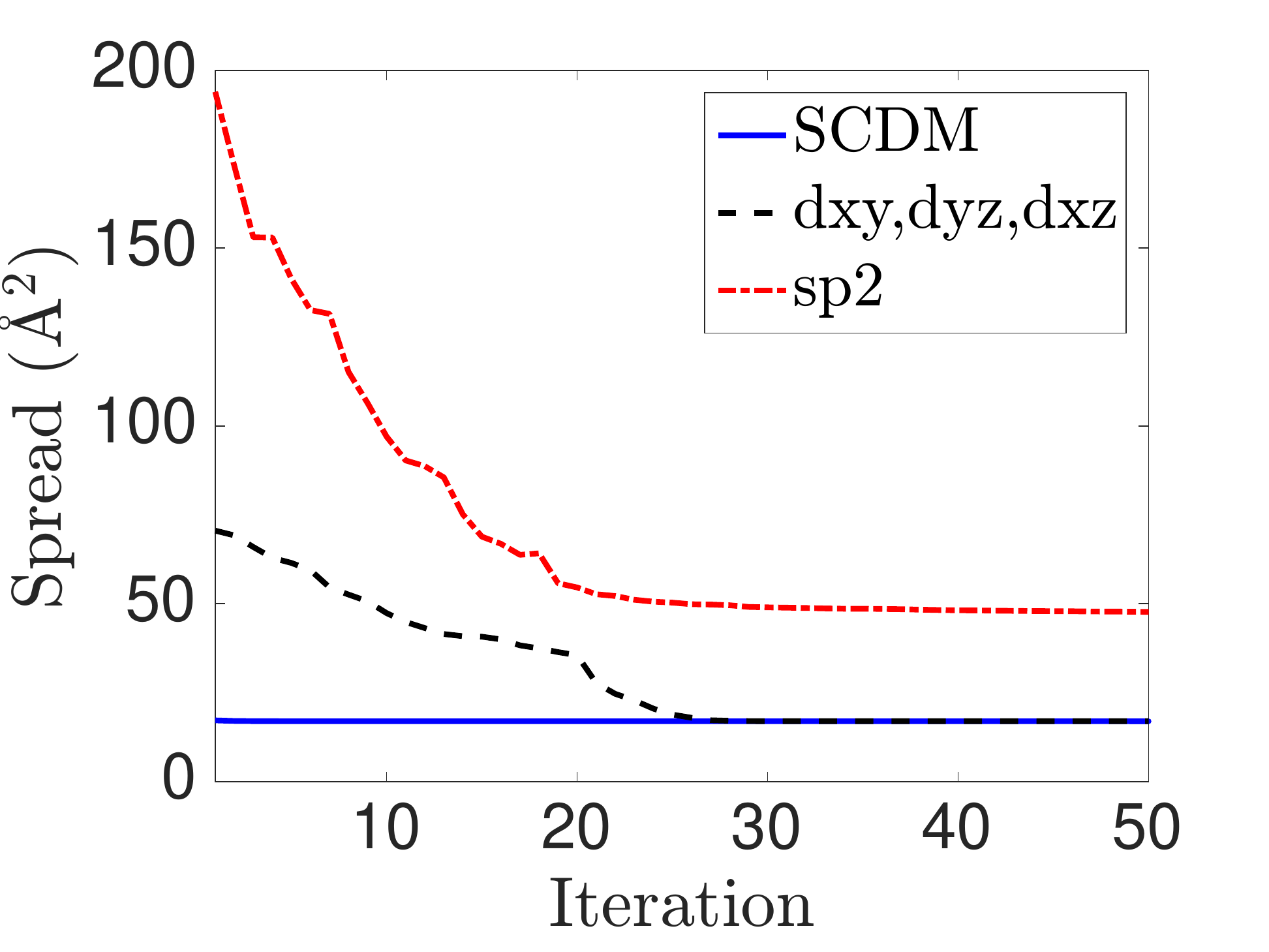}     \label{fig:cr2o3spread}}
    \subfloat[]{\includegraphics[width=0.4\columnwidth]{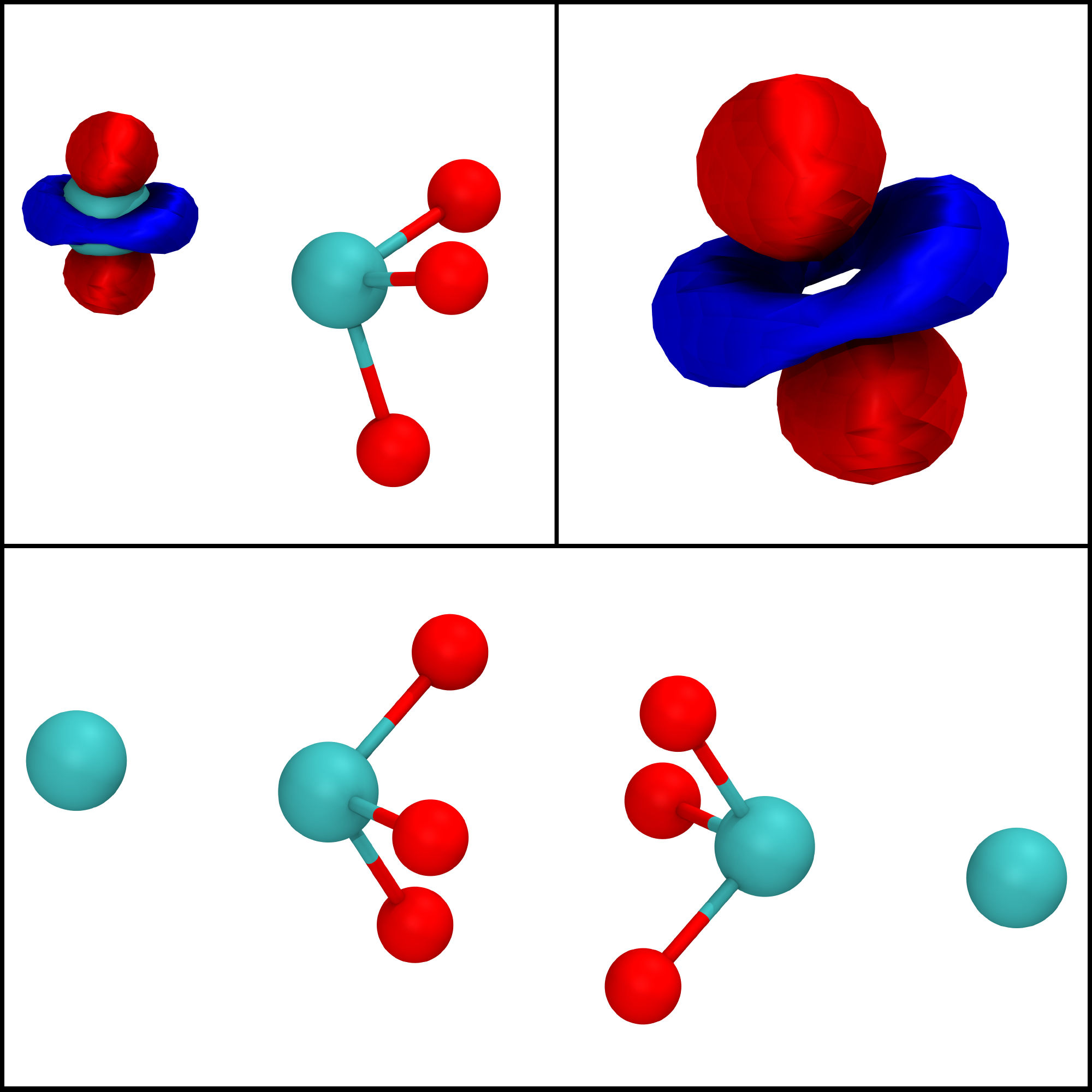} \label{fig:cr2o3struct}}
  \caption{(color online) (a) Convergence of the spread for Cr$_2$O$_3$
  starting from the initial guess of the gauge matrix from SCDM (blue
  solid line), initial guess from d orbitals (black dashed line) and
  initial guess from sp$^2$ orbitals (red dotted line).
  (b) One unit cell of Cr$_2$O$_3$, and the shape
  of a localized function obtained from SCDM (plot generated using Visual Molecular Dynamics~\cite{HUMP96}). The localized function has been plotted with and without the molecular structure to better illustrate its character and physical location. 
%  Positive and negative
%  values of the isosurface are labeled by red and blue color,
%  respectively. 
  }
    % \label{fig:cr2o3spread}

\end{figure}

Our first example is a Cr$_2$O$_3$ crystal with collinear spin polarization. Each unit cell has 92 occupied
bands and we are interested in the top $6$ valence bands, corresponding
to $3$ spin-up and $3$ spin-down $d$ orbitals for the Cr atoms. 
This is a challenging system for \texttt{Wannier90} due to the existence of multiple local
minima in the spread functional and the convergence of existing methods can
depend sensitively on the choice of the initial guess. For example, in
Fig.~\ref{fig:cr2o3spread}, when the initial guess is given by projections
corresponding to d$_{xy}$, d$_{yz}$ and d$_{xz}$ orbitals respectively,
the spread functional decreases from $70.56 \ang^2$ to $16.99 \ang^2$
within $30$ steps. In contrast, 
when the initial guess is given by sp$^2$ hybridized
orbitals the spread functional starts at $193.94 \ang^2$ and stops
decreasing around $47.13\ang^2$, indicating that the optimization
procedure is trapped at a stationary point.  On the other hand, starting
from the SCDM initial guess, the spread starts at $17.22
\ang^2$ and quickly converges to $16.98 \ang^2$.
Fig.~\ref{fig:cr2o3struct} plots the atomic configuration and 
isosurface of a localized spin-up orbital obtained from the SCDM
gauge without further Wannier optimization. The
SCDM localized orbitals clearly demonstrate $d$ orbital character without relying on a user specified initial guess.

%We now consider two examples of using our SCDM algorithm for
%disentanglement and subsequent Wannier interpolation. 
%Specifically, we
%consider eight valance and conduction bands of Silicone (corresponding
%to entangled case one for our algorithm) and seven conduction bands for
%Copper (corresponding to entangled case two for our algorithm). In
%Figure~\ref{fig:Si_and_Cu} we observe that in both cases the SCDM based
%Wannier interpolation performs well, closely matching the direct
%calculation. For the Silicone example, we used $\mu = 10$, $\sigma = 2$
%and started with 12 bands computed at each $\vk$-point. In the Copper
%case, we used $\mu = 15.5$, $\sigma = 4$, and started with 14 bands.

\begin{figure}[ht]
\centering
    \subfloat[]{\includegraphics[width=0.49\columnwidth]{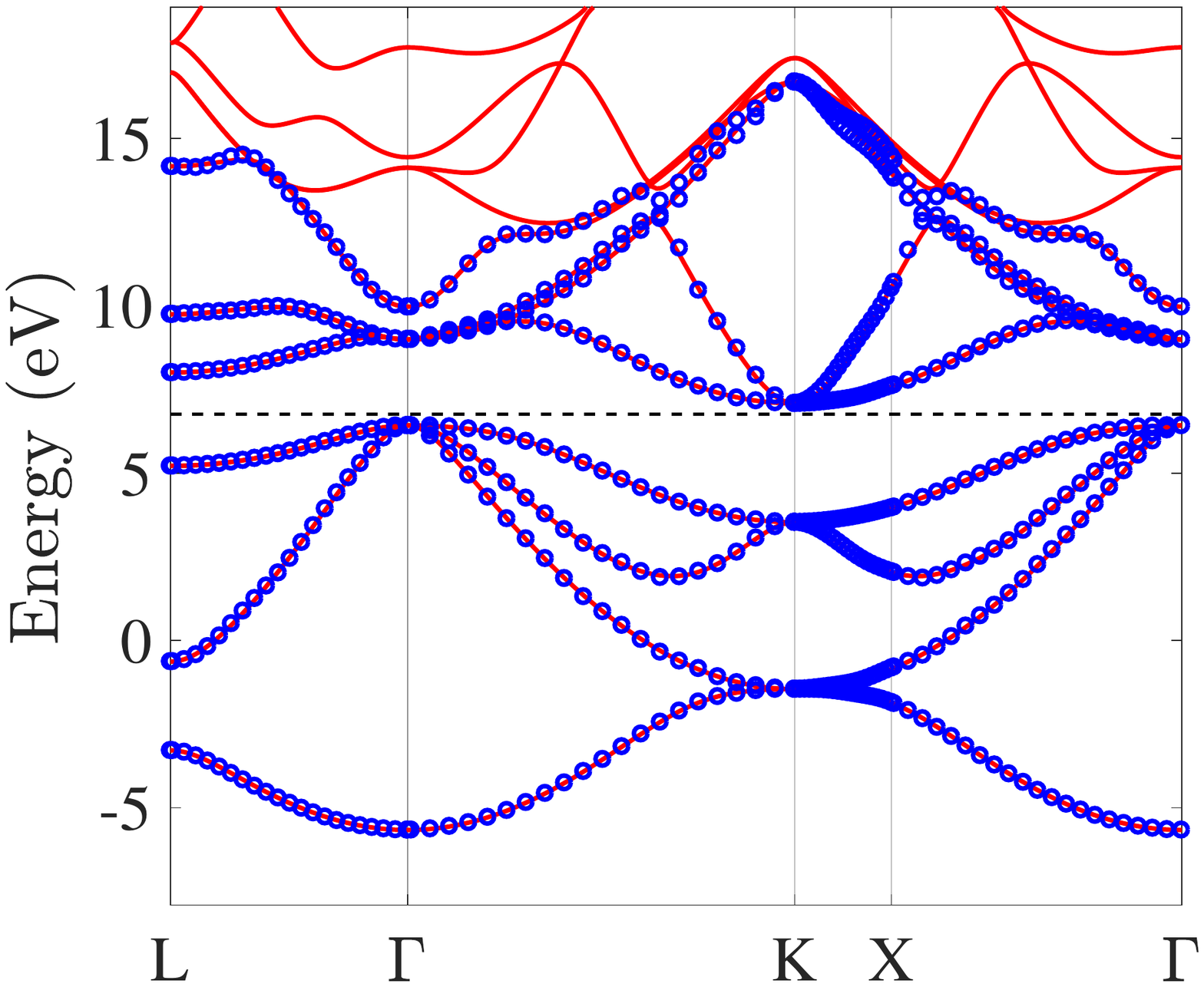} \label{fig:Si}}
    \subfloat[]{\includegraphics[width=0.48\columnwidth]{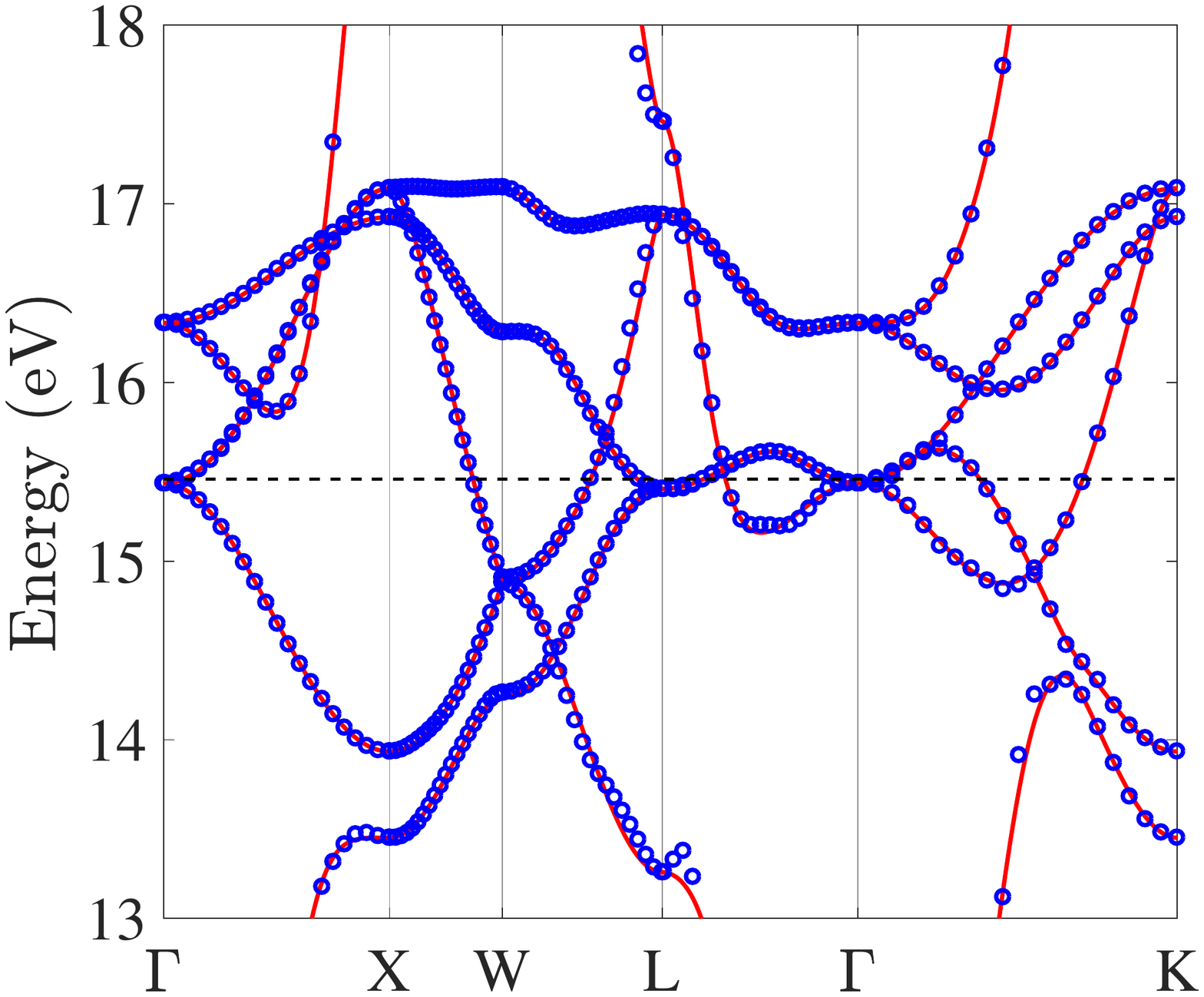} \label{fig:Cu}}
  \caption{(color online) Wannier interpolation with SCDM for the band
  structure for (a) valance and conduction bands for Si (b) bands near
  the Fermi energy for Cu. Direct calculation (red
  line) and SCDM based Wannier interpolation (blue circles).}
\end{figure}

Next, we consider two examples with entangled eigenvalues. 
Fig.~\ref{fig:Si} shows band structure
interpolation for a Si crystal with 
$8$ localized functions computed from SCDM. This corresponds 
to entangled case 1, covering both valence bands and low-lying
conduction bands. We set $\mu = 10.0$ eV, $\sigma = 2.0$ eV,
%and start with 12 bands computed at each $\vk$-point from 
and use a $10 \times 10
\times 10$ $\vk$-point grid for constructing the Wannier functions. Fig.~\ref{fig:Cu} shows the
accuracy of band structure interpolation for a Cu crystal with
$7$ localized functions. This corresponds to the entangled case 2,
covering valence bands near the Fermi energy contributed mainly from the
$d$-orbitals.
We use $\mu = 15.5$ eV, $\sigma = 4.0$ eV, and 
a $10\times 10\times 10$ $\vk$-point grid. In both cases, the SCDM
method accurately reproduces the band structure within the energy window of interest.

\begin{figure}[ht]
\centering
    \subfloat[]{\includegraphics[width=0.5\columnwidth]{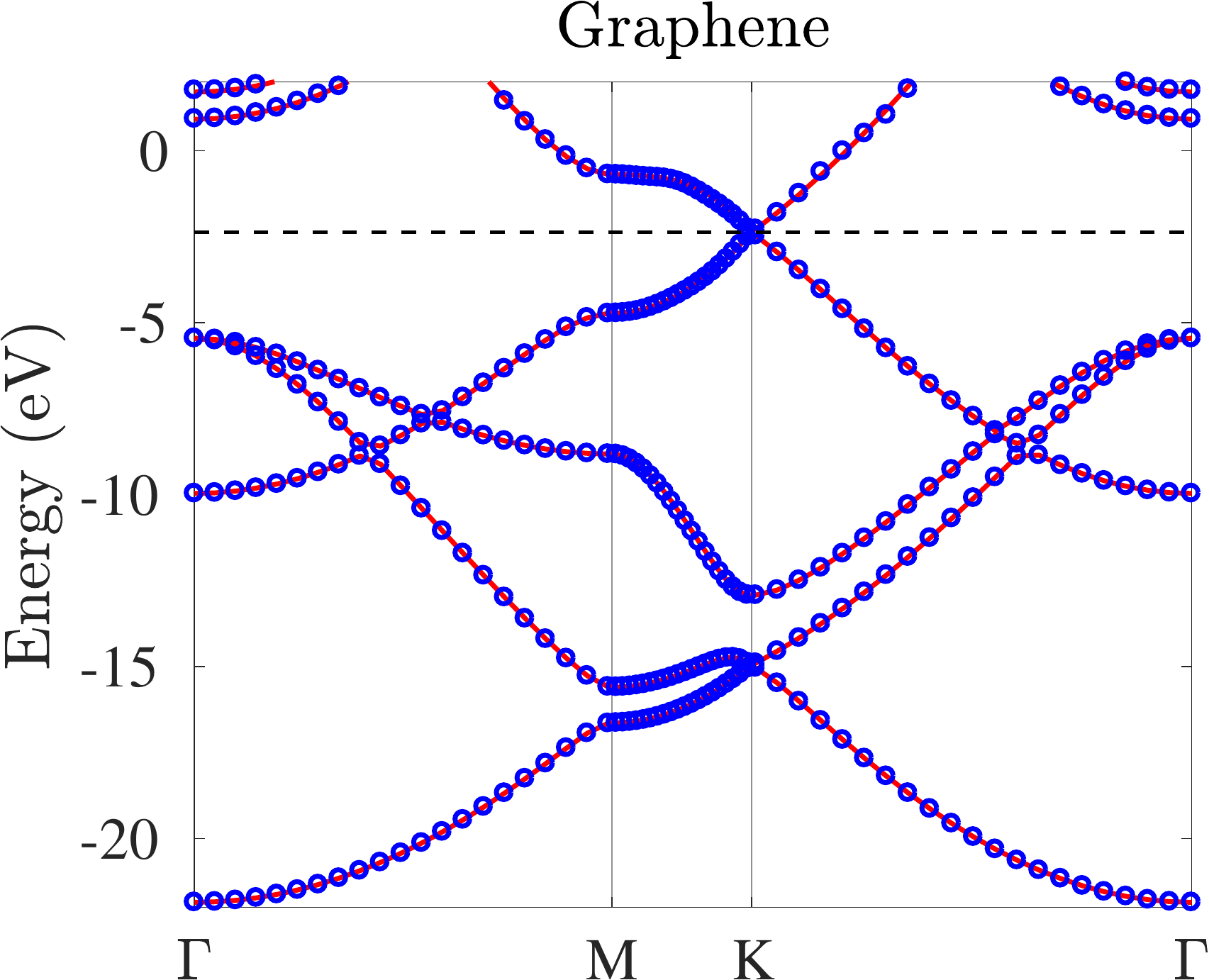}}
    \subfloat[]{\includegraphics[width=0.49\columnwidth]{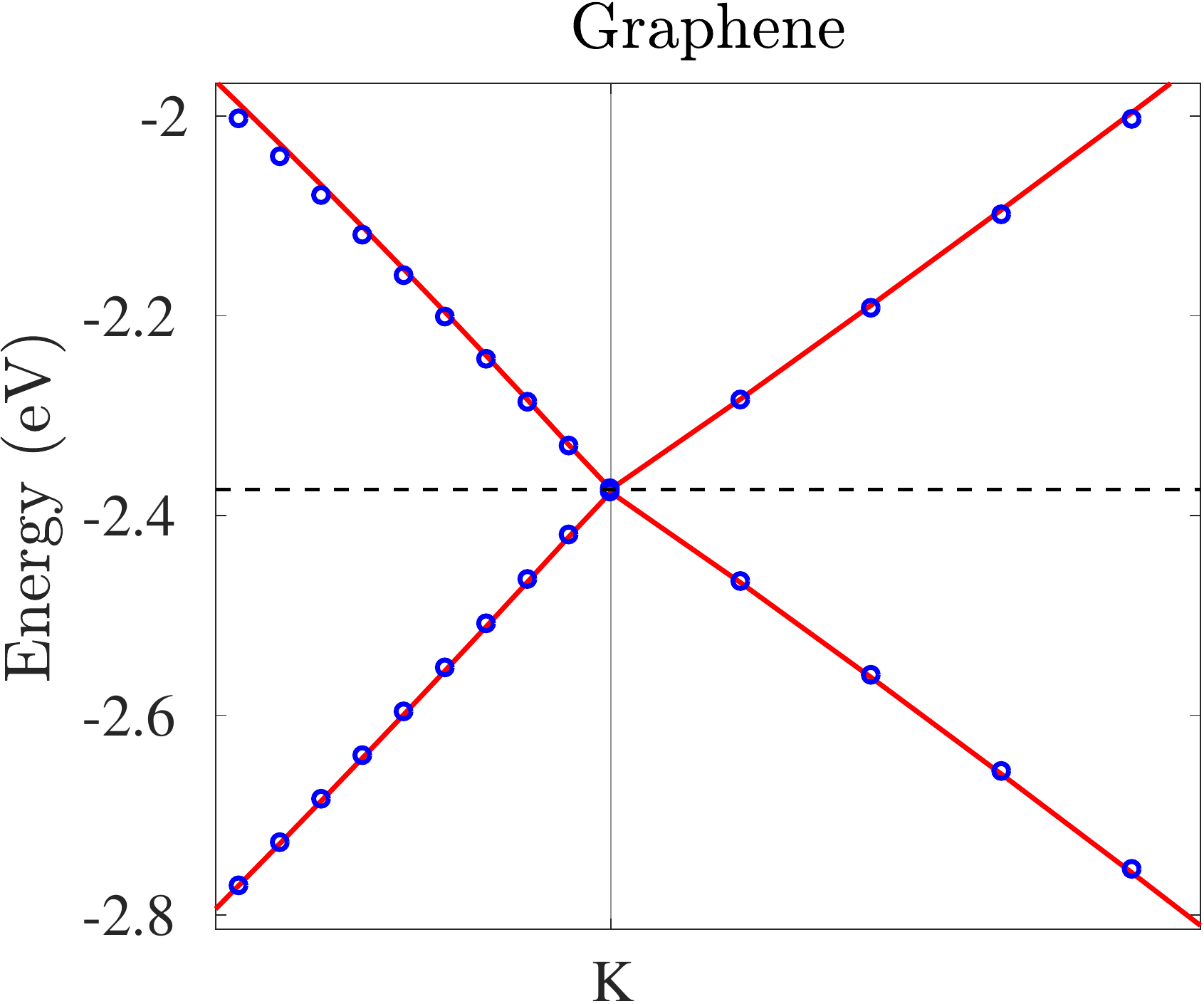}}
  \caption{(color online) Wannier interpolation with SCDM for the band structure
  for graphene (a) below the Fermi energy (b) near the Dirac point. Direct calculation (red lines) and SCDM based Wannier interpolation (blue circles).}
    \label{fig:graphene_band_kpt12}
\end{figure}

We now turn to graphene, a metallic system that is
particularly challenging for band structure interpolation due to the
linear band structure near the Dirac point.
Fig.~\ref{fig:graphene_band_kpt12} demonstrates that SCDM can accurately
interpolate the band structure of graphene even when zooming in on
the region near the Dirac point. We set $\mu=-2.5$ eV,
$\sigma=4.0$ eV, and use a $12\times 12\times 1$ $\vk$-grid for
constructing the Wannier functions.

\begin{figure}[ht]
\centering
    \subfloat[]{\includegraphics[width=0.49\columnwidth]{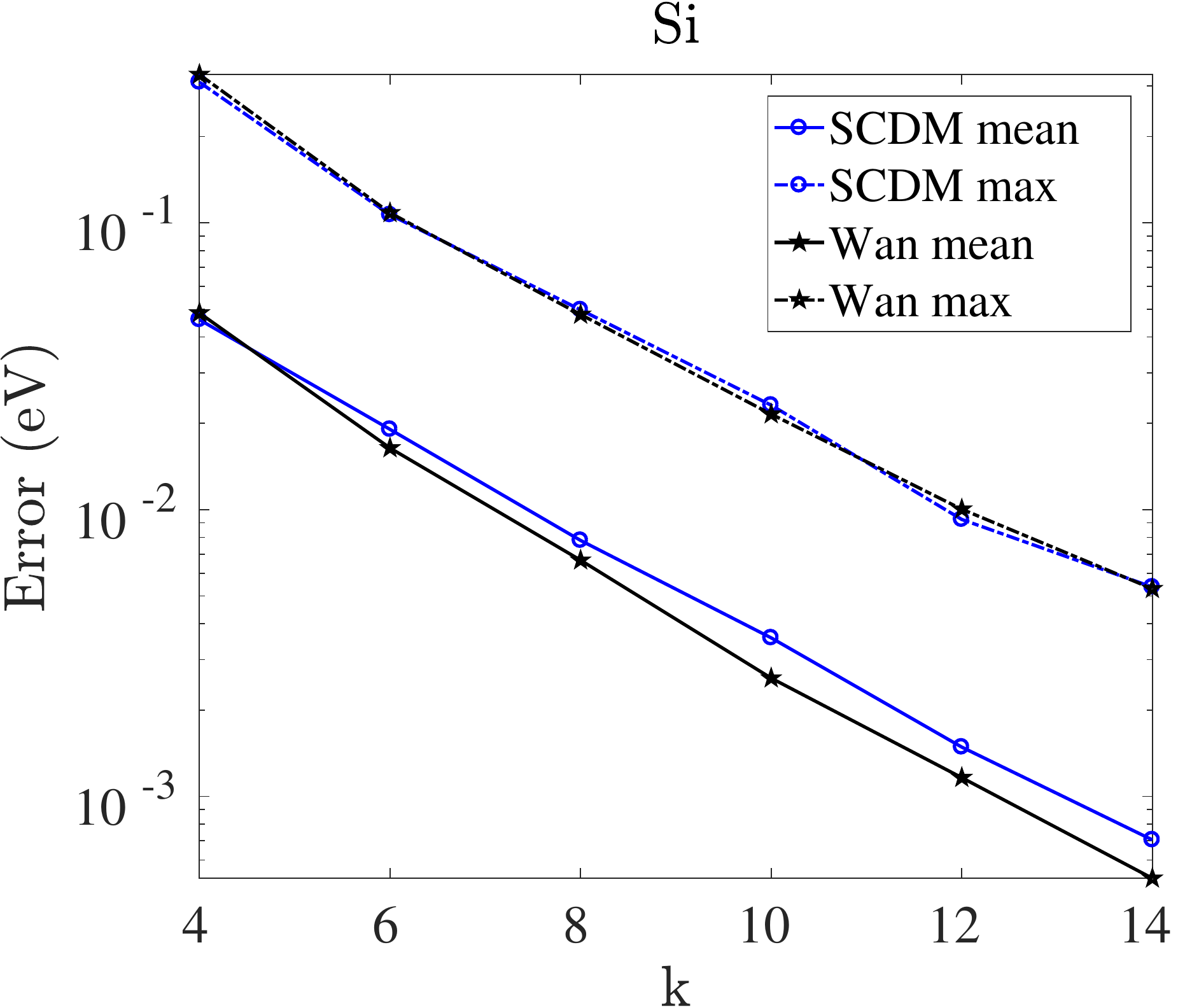}     \label{fig:si_error}}
    \subfloat[]{\includegraphics[width=0.49\columnwidth]{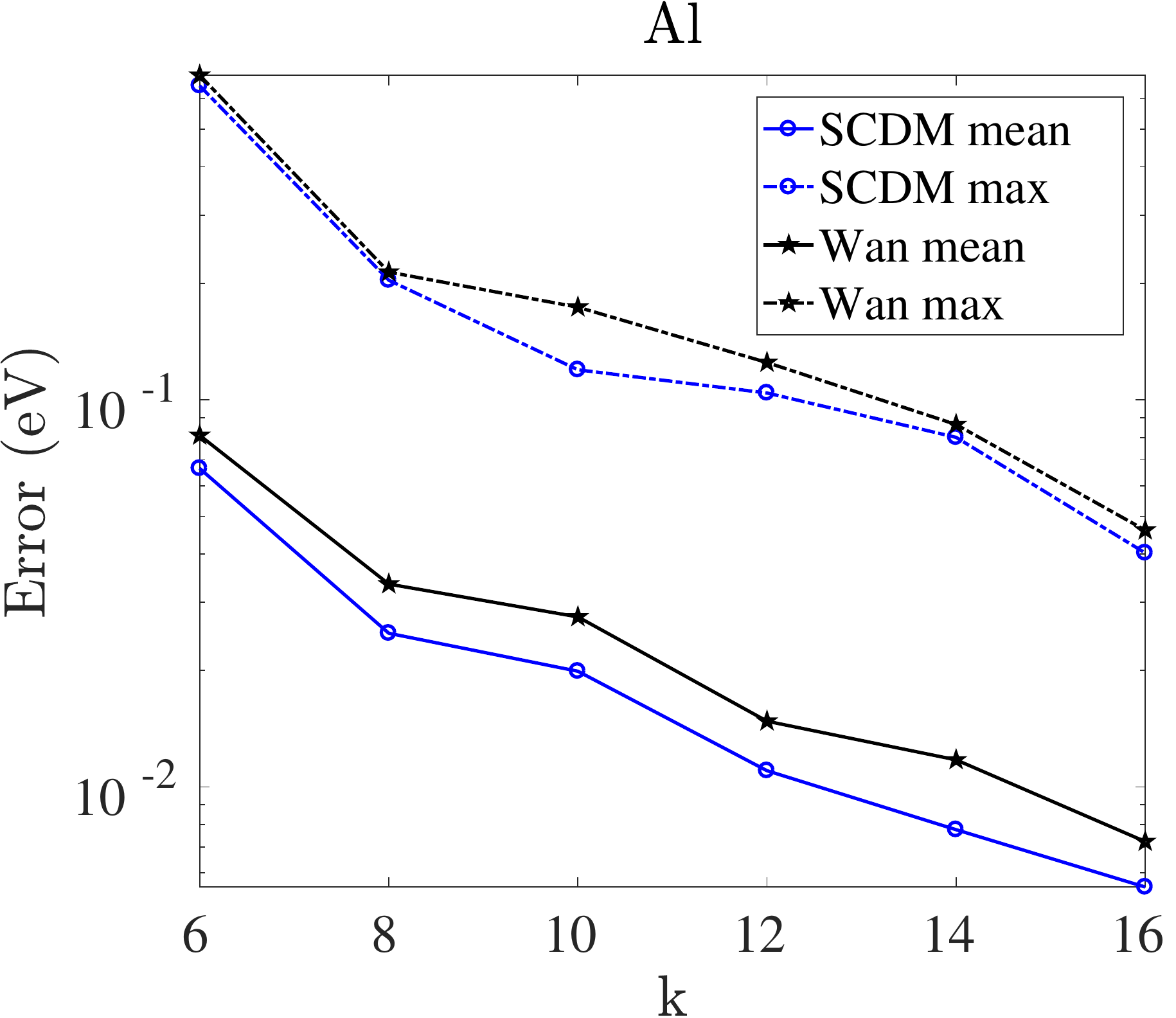}     \label{fig:al_error}}
  \caption{(color online) Convergence of the average and maximum error
  of Wannier interpolation below the Fermi energy using the SCDM gauge matrix, and
  converged Wannier gauge matrix starting from the SCDM initial guess
  for (a) silicon (b) aluminum.}
\end{figure}

Finally, we measure the convergence rate of the band structure
interpolation with respect to an increasing number
of $\vk$-points using Wannier functions obtained from SCDM and those
from a (local) minimum corresponding to the optimization objective in \texttt{Wannier90}.
Fig.~\ref{fig:si_error} reports the
absolute value of the error of the eigenvalues for the occupied
bands of Si. The chosen path through the Brillouin zone is discretized with $408$ points. 
A cubic $k\times k\times k$ grid is used and $k$ ranges from
$4$ to
$14$.  Both the average and the maximum value of the error converge
exponentially with respect to $k$. For the isolated case, the
error with the optimized gauge matrix is slightly smaller than that with
the SCDM gauge matrix. 
However, we find that visually the Wannier functions from SCDM 
with a $6\times 6\times 6$ $\vk$-grid already result in excellent
band structure.

%\begin{figure}[ht]
%  \label{fig:al_error}
%  \begin{center}
%    \includegraphics[width=0.4\textwidth]{al_error.eps}
%  \end{center}
%  \caption{(color online) Convergence of the average and maximum error of the band
%  energies below the Fermi energy using the SCDM gauge matrix, and
%  converged Wannier gauge matrix starting from the SCDM initial guess.}
%\end{figure}

\begin{figure}[ht]
\centering
    \subfloat[]{\includegraphics[width=0.5\columnwidth]{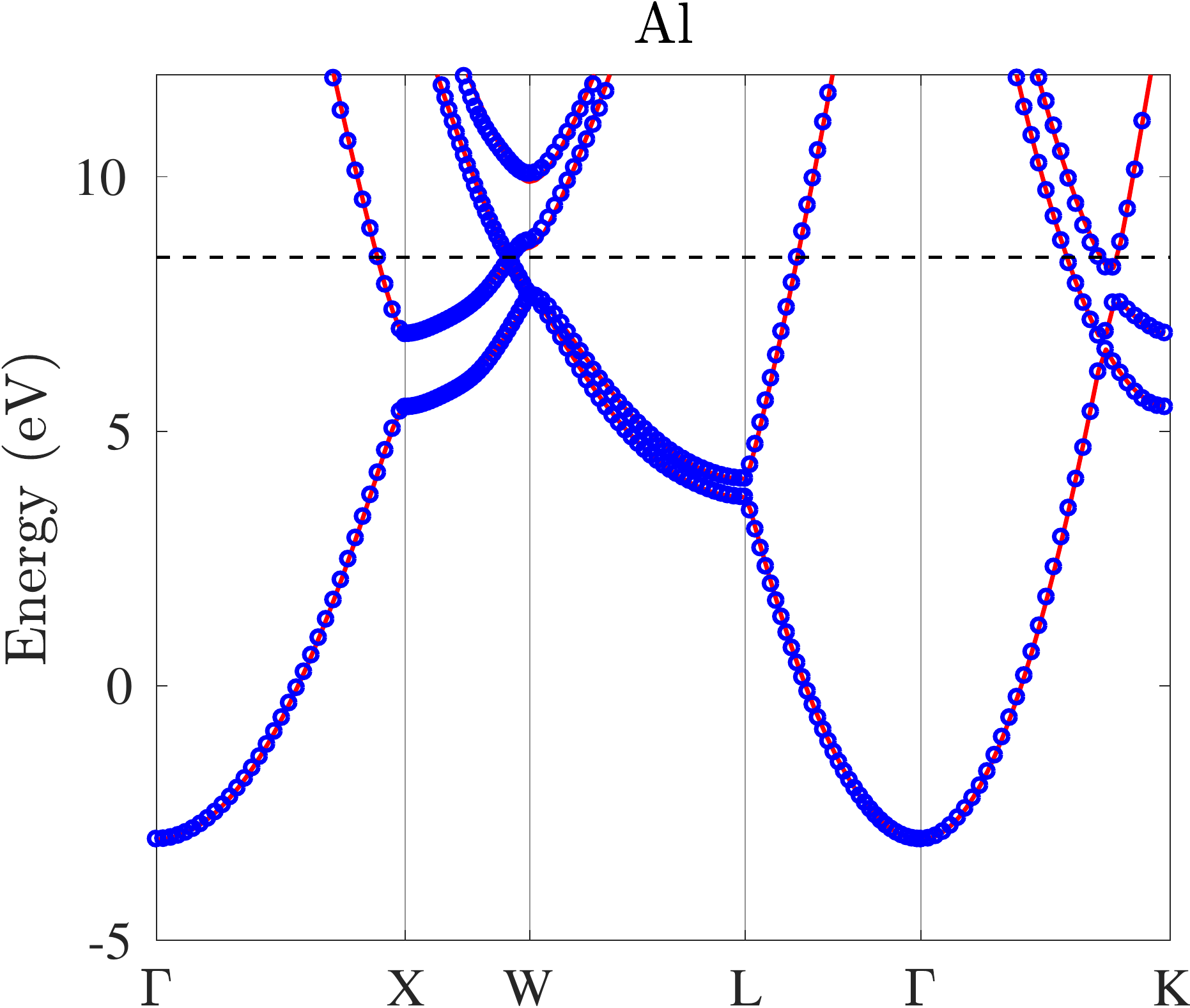} \label{fig:al_band_kpt10}}
    \subfloat[]{\includegraphics[width=0.48\columnwidth]{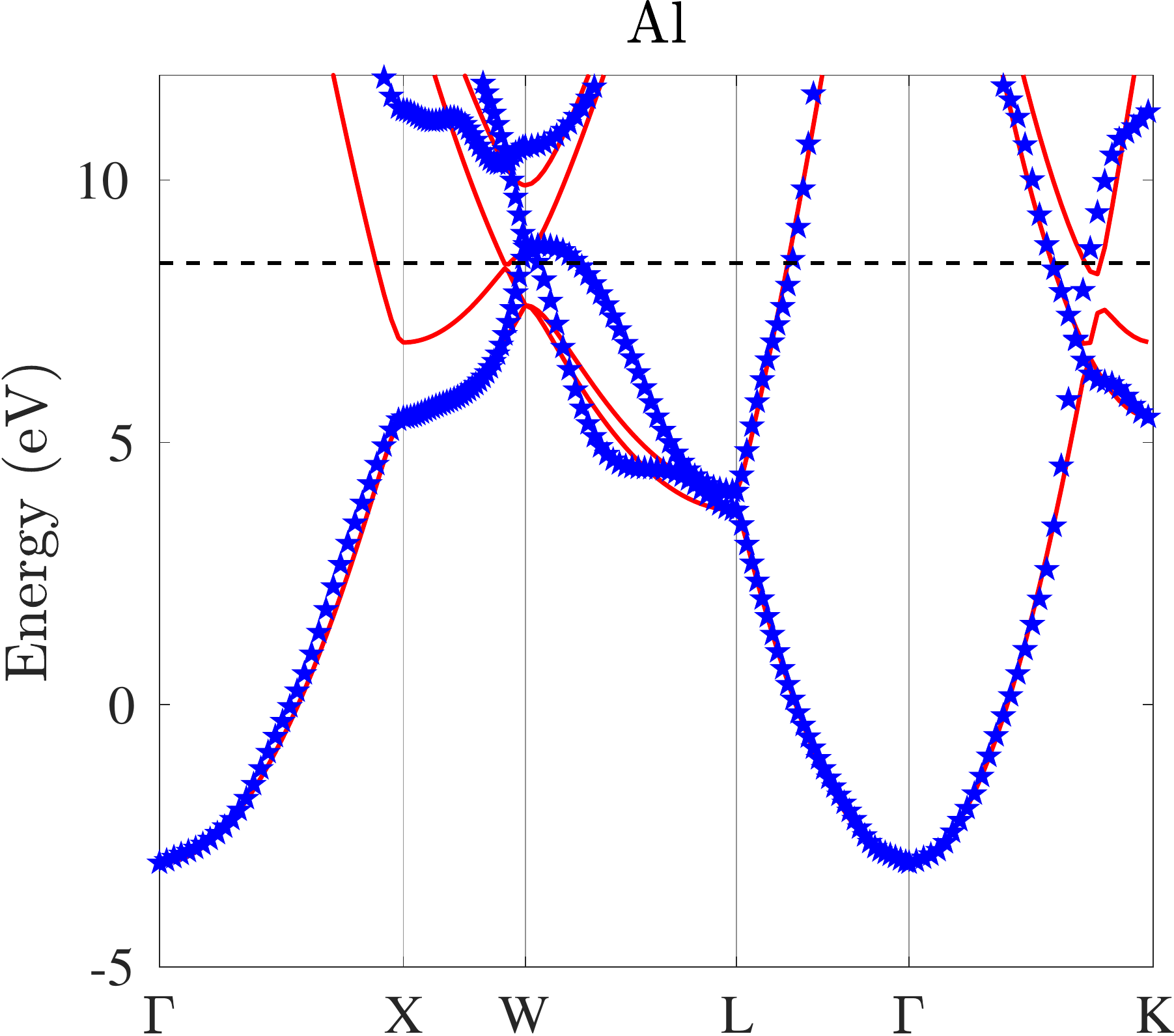} \label{fig:al_band_kpt10_wan}}
  \caption{(color online) (a) Band structure for Al around and below the Fermi energy
  (black dashed line). Direct calculation from QE (red line),
  and Wannier interpolation using SCDM gauge matrix with a $10\times
  10\times 10$ $\vk$-grid (blue circle). (b) Same calculation but using
  the Wannier gauge matrix starting from an sp$^3$ initial guess with
  $N_{b}=6,N_{w}=4$.}

\end{figure}

%
%\begin{figure}[ht]
%  \label{fig:si_band_kpt6}
%  \begin{center}
%    \includegraphics[width=0.4\textwidth]{si_band_kpt6.eps}
%  \end{center}
%  \caption{(color online) Band structure for occupied states Si. Direct calculation from QE (red line),
%  and Wannier interpolation using SCDM gauge matrix with a $6\times
%  6\times 6$ $\vk$-grid (blue circle).}
%\end{figure}
%
Fig.~\ref{fig:al_error} reports the absolute value of the error of the
eigenvalues below the Fermi energy for Al, which is a
metallic system with entangled band structure. The chosen path in the
Brillouin zone is discretized into $510$ points. 
A cubic $k\times k\times k$ grid is used and $k$ ranges from
$6$ to
$16$.  We use the $\mathrm{erfc}$ smearing with $\mu$ being the chemical
potential at $8.4$ eV, and $\sigma=4.0$ eV. We compute six bands for
each $\vk$ point and SCDM picks the leading four bands.
Even for metallic system, numerical results show
exponential convergence of the 
band structure interpolation.
Fig.~\ref{fig:al_band_kpt10} shows that  
Wannier interpolation using the SCDM gauge
matrix with a $10\times 10\times 10$ $\vk$-grid already yields excellent
band structure. In particular, using the SCDM gauge correctly
reproduces band crossings even though a relatively coarse $\vk$ grid is used. 

For this metallic system, the error of the
eigenvalues interpolated using the SCDM gauge matrix is systematically smaller than that of the
optimized gauge matrix from \texttt{Wannier90}. Therefore optimization
of the spread functional alone does not
necessarily improve the interpolation quality.
This assessment is further justified by performing Wannier interpolation with a
gauge matrix obtained by minimizing the Wannier spread functional
directly using six bands and using four orbitals that have sp$^3$
character for the initial guess. In this case the optimized spread is $12.42 \ang^2$, while
the SCDM gauge gives a larger spread of $18.38 \ang^2$. However, 
Fig.~\ref{fig:al_band_kpt10_wan} shows that the band structure obtained
using the optimized gauge with the sp$^3$ initial guess is significantly less accurate when compared to
that in Fig.~\ref{fig:al_band_kpt10} even though the same $\vk$ grid is used. In particular, the spread functional alone is not necessarily a proxy for interpolation quality. On
the other hand, the SCDM method obtains a smooth density matrix for the range of
the required band energies by construction.

% Thus, the heuristic two-stage minimization
% procedure as employed in \texttt{Wannier90} may not be the optimal
% strategy when interpolating the band structure with entangled bands. 

\section{Discussion and conclusion}
We developed a unified
method to compute Wannier functions for systems with both isolated
and entangled bands. Its simplicity\textemdash in both implementation
and reliance on few parameters\footnote{\response{Namely, essentially no
parameters in the isolated case, and two parameters $\sigma$ and $\mu$ in the entangled case.}}\textemdash makes it easy to use. Of
particular importance, our method removes the potentially sensitive
dependence of the construction of Wannier functions on an initial guess
to a nonconvex optimization (two-stage in the entangled case)
procedure. This potentially makes it easier to treat complicated
materials where the choice of a good guess may be difficult, and
convergence to local minima may hamper the construction of localized Wannier
functions. Interestingly, even though we do not seed our method with a physically informed initial guess, we
are able to recover physically interpretable Wannier functions.
Furthermore, as we have shown, the objective function of the
existing optimization procedure is not necessarily a proxy for good band interpolation. Collectively, these qualities and
observations make our new SCDM methodology attractive for the
construction of Wannier functions.

\appendix
\section{\response{Rank-revealing QR factorizations}}

Our algorithm relies heavily on a QRCP factorization, and therefore we briefly outline standard methodologies for computing these factorizations and briefly discuss the more general class of rank-revealing QR factorizations. Notably, we restrict our discussion to factorizations of matrices that are short, wide, and full row-rank as this is the setting most relevant to this manuscript. 

As outlined in Section~\ref{sec:scdm} given an $m \times n$ matrix with $m<n$ and full row-rank we seek to compute a permutation matrix $\Pi,$ an $m \times m$ orthogonal matrix $Q,$ an $m\times m$ upper triangular matrix $R_1,$ and a $m\times (n-m)$ matrix $R_2$ such that
\begin{equation}
\label{eqn:qr_appendix}
V \Pi = Q \begin{bmatrix} R_1 & R_2 \end{bmatrix}
\end{equation}
where the singular values of $R_1$ track those of $V$ as closely as possible. More specifically, we would like a factorization such that 
\[
\sigma_i\left(R_1\right) \geq \frac{\sigma_i\left(V\right)}{g(n,k)}
\] 
for some function $g$ of $n$ and $k.$ The specific form of $g(n,k)$ depends on the algorithm used, and there has been significant work on developing algorithms to achieve $g(n,k)$ growing as slowly as possible in $n$ and $k.$ We direct the reader to references~\cite{chandrasekaran1994rank} and~\cite{GuEisenstat1996} for further details.

While more recent rank-revealing algorithms may be necessary for certain problems, often viewed as pathological worst case examples, the most widely used rank-revealing QR factorization is a QR factorization with column pivoting due to Golub and Businger~\cite{businger1965linear}. While formally this algorithm exhibits a rather weak form of $g(n,k),$ its implementation in LAPACK~\cite{LAPACK} and strong practical performance has driven its use. This practical performance holds true remarkably robustly in our setting and for all the problems we have considered.

The underlying algorithm\footnote{A textbook presentation may be found
in Golub and Van Loan~\cite{GolubVan2013}} is encapsulated by a simple
heuristic strategy for picking $\Pi.$ This is outlined in
Algorithm~\ref{alg:qr}, and can be colloquially summarized as greedily
picking columns at each step that look the least like those already
selected. At the conclusion of Algorithm~\ref{alg:qr} the first $m$
entries of $\pi$ yield the information we require about which $m$
columns were selected as pivots during the course of the algorithm, and
we have omitted any reference to $Q$ or $\begin{bmatrix} R_1 & R_2
\end{bmatrix}$ for simplicity.  Importantly, we emphasize that
Algorithm~\ref{alg:qr} is a conceptual presentation of the algorithm. In practice an implementation would differ significantly from this description.  

\begin{algorithm}
\caption{A conceptual description of the Golub and Businger QR factorization with column-pivoting specialized to the SCDM setting.}
\label{alg:qr}
\textbf{Input:} $V\in\mathbb{C}^{m\times n}$ with $m<n$ and full row-rank
\begin{algorithmic}[1]
\State Initialize $\pi(i) = i$ for $i=1,2,\ldots,n$
\For{$k=1,2,\ldots,m$}
\State Set $$ j = \argmax_{i=k,\ldots,n}\|V(k:m,i)\|_2 $$
\State Swap $\pi(k)$ and $\pi(j),$ and $V(:,k)$ and $V(:,j)$
\State Construct a unitary matrix $H^{(k)}$, a so-called Householder
reflector such that
$$H^{(k)}V(k:m,k) = \pm \|V(k:m,k)\|_2e_1$$
\State Set $V(k:m,k:n) = H^{(k)}V(k:m,k:n)$
\EndFor
\State \textbf{Return:} $\pi$
\end{algorithmic}
\end{algorithm}

% First, the column with largest norm is picked to be the first ``pivot'' and permuted to the front. Then, a single step of a standard QR factorization is completed via application of a Householder reflector. Now, to select the second pivot and determine the second column of $\Pi$ we simply take the column that has largest remaining norm after its component in the direction of the already selected column has been projected out. Given the dimensionality of our setting, this process simply repeats until we have selected $m$ columns and at step $k$ a column is chosen based on its distinctiveness from the prior $k-1$ selected columns. 

\section{\response{One dimensional model problem}}
\label{a:onedim}
\begin{figure}[ht]
\centering
    \includegraphics[width=.8\columnwidth]{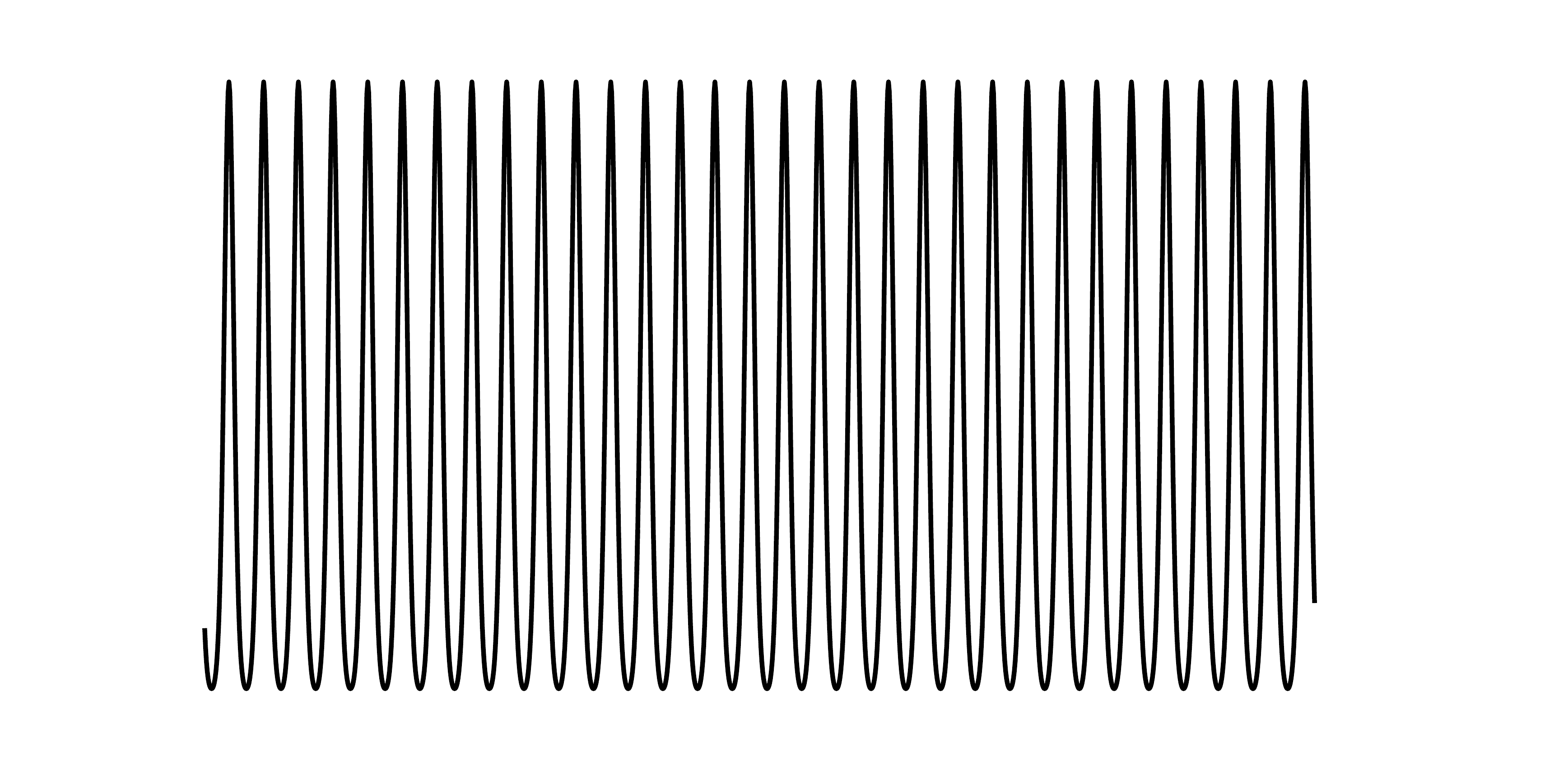}
  \caption{The potential function for our one dimensional model problem.}
  \label{fig:model_potential}
\end{figure}

\begin{figure}[ht]
\centering
    \subfloat[]{\includegraphics[width=0.3\columnwidth]{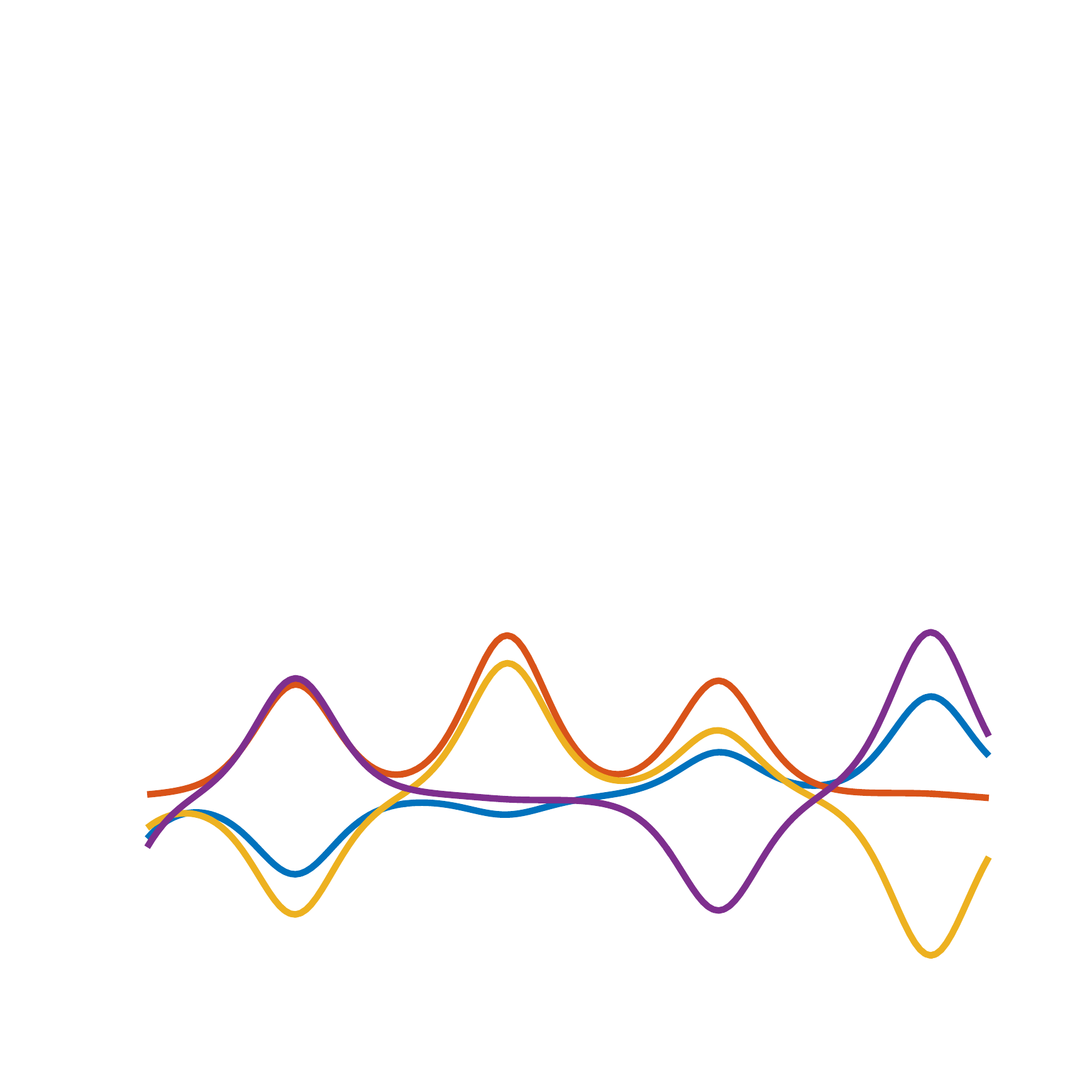}}\qquad
    \subfloat[]{\includegraphics[width=0.3\columnwidth]{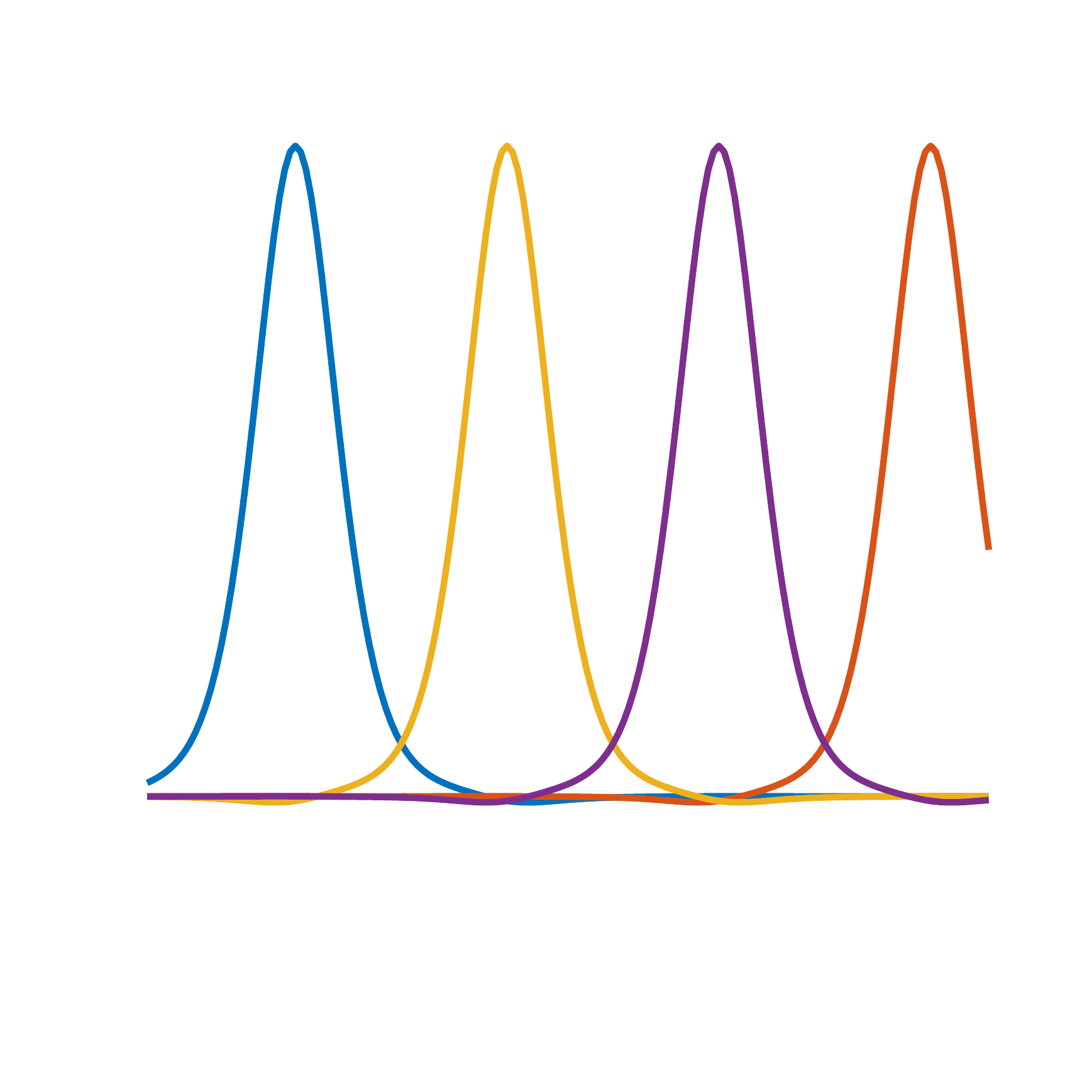}}\\
    \subfloat[]{\includegraphics[width=0.3\columnwidth]{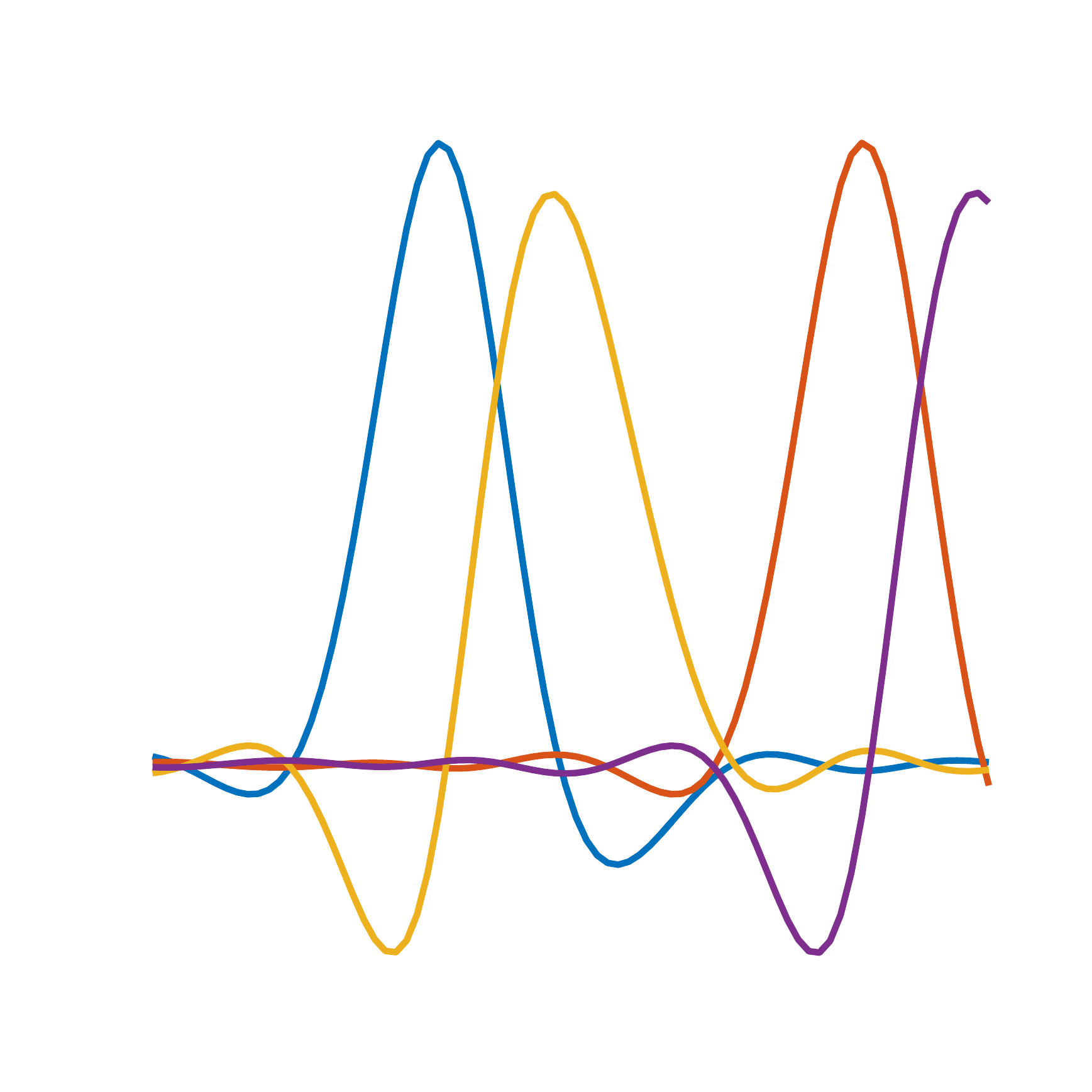}}\qquad
    \subfloat[]{\includegraphics[width=0.3\columnwidth]{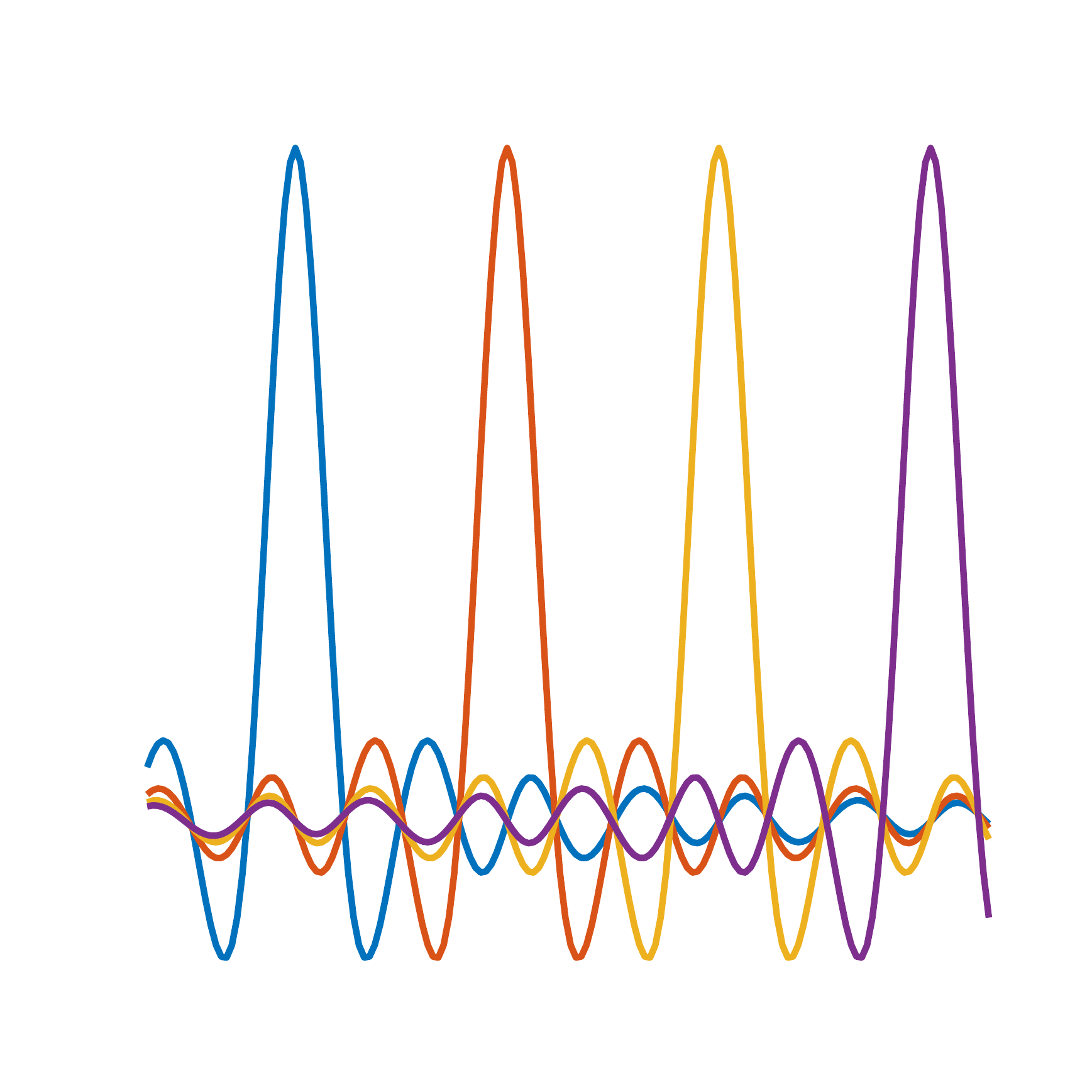}}
  \caption{For a simple one dimensional model problem, we plot the
  eigenfunctions (a) and Wannier functions for (b) the isolated case,
  (c) entangled case 1, and (d) entangled case 2. Here, only a portion
  of the computational domain (horizontal axis) is shown to more clearly
  illustrate the local structure of the Wannier functions. 
  }
  \label{fig:model_more}
\end{figure}

The one dimensional model problem consists of a Hamiltonian operator based on the one-dimensional periodic potential plotted in Figure~\ref{fig:model_potential} and discretized on 1280 points. Figure~\ref{fig:model_more} illustrates additional localized functions for each of the three cases. Here we observe that their behavior matches that of the examples shown in the main text. Here, the size of the problem allowed us to computed all 1280 eigenfunctions and only consider the ones of interest. In the first isolated case we set $\mu_c = 0.2$ and $\sigma = 0.5,$ and in the second case we set $\mu_c = 0.15$ and $\sigma = 2.$ These plots may be reproduced using the included data and files in the code repository available at \url{https://github.com/asdamle/SCDM}.

\section*{Acknowledgments}
The authors thank Stefano Baroni, Eric Canc\`es, Roberto Car, Sinisa
Coh, Wibe de Jong, Antoine Levitt, Jianfeng Lu, Nicola Marzari, Lukas
M\"uchler and Lexing Ying for useful discussions; Sinisa Coh
for providing the atomic structure for the Cr$_2$O$_3$ example; and the
anonymous referees for their helpful suggestions. 

\FloatBarrier

\bibliographystyle{siamplain}
\bibliography{wannier}
\label{LastPage}
\end{document}